\documentstyle[12pt]{article}
\newcommand{\ju}[4]{\mbox{$
  \left(\begin{array}{cc}{#1} & {#2}\\{#3} &{#4}  \end{array}\right)$}}

\newcommand{\la}{\lambda}

\newcommand{\pa}{\partial}

\newcommand{\be}{\begin{equation}}
\newcommand{\ee}{\end{equation}}

\newcommand{\La}{\Lambda}

\title{\Large \bf Generalized  r-matrix structure and
 algebro-geometric solution for integrable systems}
\author{Zhijun Qiao$^{a,b,c}$\\
\small $^a$Institute of Mathematics, Fudan University, Shanghai
200433, P. R. China\\
\small $^b$Theoretical Division, Los Alamos National Laboratory, 
Los Alamos,  NM 87545, USA\\
\small
$^c$Fachbereich 17, Mathematik, Universit\"{a}t-GH Kassel,\\
\small Heinrich-Plett-Str. 40, D-34109 Kassel, Germany\\
\small Email \ qiao@lanl.gov \ \ \  zhijun@uni-kassel.de}
\date{\small \sf This paper has been published in Rev. Math. Phys, 13(2001), 545-586}
\begin{document}
\maketitle                    
\begin{abstract}
The purpose of this paper is to construct
 a generalized $r$-matrix structure of
 finite dimensional systems and an
approach to obtain the algebro-geometric solutions of
integrable nonlinear evolution equations (NLEEs).
Our starting point is a generalized Lax matrix instead of
usual Lax pair. The generalized $r$-matrix structure
and Hamiltonian functions are
presented on the basis of fundamental Poisson bracket.
 It can be clearly seen that various nonlinear
 constrained (c-) and restricted
 (r-) systems,
such as the c-AKNS, c-MKdV, c-Toda, r-Toda, c-Levi, etc, are derived from
the reduction of this structure.
All these nonlinear systems have
{\it r}-matrices, and are completely integrable in Liouville's sense.
Furthermore, our generalized structure is developed to become
an approach to obtain the algebro-geometric solutions
of integrable NLEEs.
Finally, the two typical examples
are considered to illustrate this approach:
the infinite or periodic Toda lattice equation
and the AKNS equation with the condition of
decay at infinity or periodic boundary.
\end{abstract}

$ $

\section{\bf Introduction}
Completely integrable systems
are widespreadly applied in field theory, fluid mechanics,
nonlinear optics and other fields of nonlinear sciences.
The new development
of integrability theory can be roughly divided into three stages. The
first one was the direct use of
Lax equations for some systems such as the Calogero-Moser
system \cite{MJ} and the Euler rigid equation \cite{VM},
which were allowed for integration.
These systems were not amenable to the classical techniques of integrating
Hamiltonian equations of motion. The second
one was the so-called `algebraization', i.e., the tools of Lie algebras,
Kac-Moody algebras
were used to sysmetically construct a large class of soliton equations
and integrable systems
\cite{AM,RAS,S}, and simultaneously present the Lax representations
 of soliton equations, Hamiltonian structures. The third one
is being developed and witnessed through the use of nonlinearization
technique \cite{C1} to generate finite-dimensional
integrable systems. These systems can be
the Bargmann system, the C. Neumann system \cite{CG1,Qiao1},
the higher-order constrained flows or symmetric constrained flows
\cite{AW1,AW2},
and the stationary flows of
soliton equations \cite{Wo}. Indeed,
with the help of this method,
many new completely integrable systems
were successively found \cite{CG1,Qiao1}. In this way,
each integrable system is generated through making
nonlinearized procedure for a concrete spectral problem
or Lax pair, and it has its own characteristic property.
Then a natural question is whether or not there is a unified
structure such that it can contain those concrete
integrable systems?
Recently, the study of $r$-matrix for nonlinear
integrable systems brings a great hope to solving this problem.

Semenov-Tian-Shansky
ever gave the definition of $r$-matrix \cite{S}, and used the $r$-matrix
to construct Lie algebra and new Poisson bracket \cite{RAS1}
in a given Lie algebra and corresponding coadjoint orbit.
The main idea of Semenov-Tian-Shansky and Reyman was how to
obtain the new Poisson bracket from a given $r$-matrix and an element
of Lie algebra. Here our thought is how to present $r$-matrix
structure from a given Lax matrix
and the standard Poisson bracket:
$$ L,\ \{\cdot,\cdot\} \stackrel{?}{\Longrightarrow} \ r-matrix. $$

In the present paper, we give a sure answer for the above
question. We
propose an approach to generate finite dimensional
integrable systems by beginning with the so-called generalized Lax matrix
instead of usual Lax pair.
Another main result of this paper is to deal with the
algebro-geometric solutions of integrable
nonlinear evolution equations (NLEEs).
It is well-known that the ideal aim for nonlinear equations
is to obtain their explicit solutions. According to
the nonlinearization method, solutions of
integrable NLEEs can have the parametric
representations \cite{C5} or
involutive representations \cite{C6}, and also have numeric
representations in the discrete case \cite{RCW}. However,
these representations of solutions are not given in an explicit form.
Thus, an open question is how to obtain their explicit
forms. In the paper we would like to give solutions of
integrable NLEEs in the form of algebro-geometric
$\Theta$-functions.

The algebro-geometric solutions for some
soliton equations with the periodic boundary
value problems were known since the works of Lax \cite{L},
Dubrovin, Mateev and Novikov \cite{DMN}.
Similar results for the periodic Toda case were obtained
slightly later by Date and Tanaka \cite{DT}.
Afterwards, the relations between commutative rings and ordinary
linear periodic differential operators and between algebraic curves and
nonlinear periodic difference equations were discussed
by Krichever \cite{Kr}. The technique they adopted
is the Bloch eigenfunctions, the spectral theory
of linear periodic operators, and some analysis tools on Riemann surfaces.

In the first example of this paper, we give the algebro-geometric
solutions of the Toda lattice equation in the
infinite or periodic case. Our method
is a constraint approach connecting finite dimensional integrable
systems with integrable NLEEs instead of usual spectral
techniques and Bloch eigenfunctions which are often available to
the periodic boundary problems. The results with the periodic
boundary conditions are included in ours. In the second example,
we consider the well-known
AKNS equation.
The Ablowitz-Kaup-Newell-Segur (AKNS)
equations are a very important hierarchy \cite{AKNS1} of
NLEEs in soliton theory. It can turn
out that the KdV, MKdV, NLS, sine-Gordon, sinh-Gordon equations etc.
All these equations are solvable by the inverse scattering transform
(IST) \cite{GGKM1}, and usually have $N$-soliton solutions \cite{AS}.
But the algebro-geometric solutions of the AKNS equations
have not been obtained since then.
We shall solve this problem
by using our constraint procedure. The considered AKNS equation
is under the case of decay at infinity or periodic boundary condition.

The whole paper is organized as follows.
We first introduce a generalized Lax matrix in the next
section, then construct
a generalized {\it r}-matrix structure and a generalized set of involutive
Hamiltonian functions in section 3. All those
Hamiltonian systems have Lax matrices,
{\it r}-matrices, and are therefore completely integrable in Liouville's sense.
In section 4 it can be seen that various 
constrained (c-) and restricted (r-) integrable flows,
such as the c-AKNS, c-MKdV, c-Toda, r-Toda, c-WKI, c-Levi, etc,
can be derived from
the reductions of this structure. Moreover, the following
interesting facts are given in sections 5, 6, 7, respectively:

  --  Several pairs of different integrable systems share the same
        {\it r}-matrices with the good
       property of being non-dynamical (i.e. constant). In particular,
       a discrete and a continuous dynamical system possess the common Lax matrix, $r$-matrix, and even completely same involutive set. Additionally,
       on a symplectic submanifold integrability of
       the restricted Hamiltonian flow (for continuous case) and
       symplectic map (for discrete case) are described by introducing
       the Dirac-Poisson bracket. They also have the same $r$-matrix
       but being dynamical.

  --  A pair of constrained integrable systems, produced by two gauge
       equivalent spectral problems, posesses different
       {\it r}-matrices being non-dynamical.

  --  New integrable systems are generated through choosing
       new {\it r}-matrices from our structure, and the associated
       spectral problems are also new.\\
In the last section, as a development of the generalized structure,
through considering the relation lifting finite dimensional
system to infinite dimensional system and using the algebro-geometric
tools we present an approach for obtaining the algebro-geometric
solution of integrable NLEEs. 
To illustrate the procedure we take 
the periodic or infinite Toda lattice equation 
and the AKNS equation with the condition of
decay at infinity or periodic boundary as the examples.

     Before displaying our main results, let us first give some
     necessary notation:
    $dp\wedge dq$ stands for the standard symplectic structure
in Euclidean space $R^{2N}=\{(p,q)|p=(p_1,\ldots,p_N), q=(q_1,\ldots,q_N)\}$;
$<\cdot,\cdot>$ is the standard inner product in $R^N$;
in the symplectic space $(R^{2N},dp\wedge dq)$
the Poisson bracket of two Hamiltonian functions
$F,G$ is defined by \cite{AV}
\begin{equation}
 \{F,G\}=\sum_{i=1}^N(\frac{\partial F}{\partial q_i}\frac{\partial G}{\partial p_i}
 -\frac{\partial F}{\partial p_i}\frac{\partial G}{\partial  q_i})=<\frac{\partial F}{\partial q},\frac{\partial G}{\partial p}>
 -<\frac{\partial F}{\partial p},\frac{\partial G}{\partial  q}>;
\end{equation}
$I$ and $\otimes$ stand for the $2\times2$ unit matrix and
the tensor product of matrix, respectively;
$\lambda _1,...,\lambda _N$ are $N$ arbitrarily given distinct constants;
 $\lambda,\mu$ are the two different spectral parameters;
$\Lambda=diag(\lambda _1,...,\lambda _N),\ I_0=<q,q>,\ J_0=<p,q>, K_0=<p,p>$,
$\ I_1=<\Lambda p,p><\Lambda q,q>,\ J_1=<\Lambda p,q>,\
 a_0, \ a_1=const.$.
Denote all infinitely times differentiable functions on real field $R$ by
$C^{\infty}(R)$.

\section{A generalized Lax matrix}
\setcounter{equation}{0}

Consider the following matrix (called Lax matrix)
\begin{equation}
L(\lambda)=\left(\begin{array}{lr}
A(\lambda) & B(\lambda)\\
C(\lambda) & -A(\lambda)
\end{array}\right)
\end{equation}
where
\begin{eqnarray}
A(\lambda)&=&a_{-2}(I_1,J_1)\lambda^{-2}+a_{-1}(J_0)\lambda^{-1} + a_0 + a_1\lambda+
           \sum_{j=1}^{N}\frac{p_{j}q_{j}}{\lambda-\lambda_{j}}, \\
B(\lambda)&=&b_{-1}(I_0,J_0)\lambda^{-1}+b_0(J_0)-\sum_{j=1}^{N}\frac{q_{j}^{2}}{\lambda-\lambda_{j}},\\
C(\lambda)&=&c_{-1}(J_0,K_0)\lambda^{-1}+c_0(J_0)+\sum_{j=1}^{N}\frac{p_{j}^{2}}{\lambda-\lambda_{j}}.
\end{eqnarray}
with some undetermined functions $a_{-2}, \ a_{-1}, \ b_{-1}, \ c_{-1},
\ b_0,\ c_0 \in C^{\infty}(R)$.

Now, in order to produce finite dimensional integrable systems directly from the Lax matrix
(2.1), we need an inevitable {\bf Assumption (A)}:
$\{A(\lambda),A(\mu)\}, \\ \{A(\lambda), B(\mu)\}, \{A(\lambda), C(\mu)\},
\{B(\lambda), B(\mu)\}, \{B(\lambda), C(\mu)\}$, and $\{C(\lambda), C(\mu)\}$
are expressed as some linear combinations of
$A(\lambda),A(\mu), B(\lambda), B(\mu), C(\lambda), C(\mu)$ with
the cofficients in $C^{\infty}(R)$. Then we have

     {\bf Lemma 2.1} \quad { \it Under {\bf Assumption (A)},
$L(\lambda)$ only contains the following cases:

1.  If $a_{-2}\not = const.$, $a_0=b_0=c_0=a_1=0,\ a_{-1}=-J_0,\ b_{-1}=I_0,$
and $c_{-1}=-K_0,$ then $a_{-2}$ satisfies the relation
$I_1=(J_1+a_{-2})^2+f(a_{-2})$;
if $a_{-2}=const.\not = 0$ and $a_0=b_0=c_0=a_1=0,$ then
 $a_{-1}=-J_0, \  b_{-1}=I_0, \ c_{-1}=-K_0$, or
$\ a_{-1}=const.,
\ b_{-1}=I_0+f_1(J_0),$ $c_{-1}=-K_0+g_1(J_0)$, where $f_1,\ g_1$
satisfy the relation
$f_1g_1=-J_0^2-2a_{-1}J_0+const.$.

2.   $a_{-2}=a_{-1}=b_{-1}=c_{-1}=b_0=c_0=a_1=0,$ and $a_0=const.$

3.   $a_{-2}=b_{-1}=a_0=b_0=c_0=a_1=0, \ c_{-1}=-K_0,$ and
    $a_{-1}$ satisfies $\frac{da_{-1}}{dJ_0}\ne 0.$

4.   $a_{-2}=a_{-1}=b_{-1}=c_{-1}=b_0=a_1=0, \ a_0=const.,$ and $c_0\ne 0.$

5.   $a_{-2}=c_{-1}=b_0=a_1=0, \ a_{-1}, \ a_0=const., \ b_{-1}=I_0+g(J_0)$,
   and  $c_0$ satisfies $\frac{d}{dJ_0}(c_0g(J_0))=-2a_0$.

6.   $a_{-2}=c_{-1}=a_0=b_0=c_0=a_1=0, \ a_{-1}=J_0+const.,$ and $b_{-1}=I_0.$

7.   If $a_{-2}=a_0=b_0=c_0=a_1=0$, then
    there are the following five subcases:

\hspace{0.55cm}{\small  (7.1)   $a_{-1}=const.,\ c_{-1}=-K_0+f_2(J_0),$ and $
    b_{-1}=I_0+g_2(J_0);$

\hspace{0.42cm}  (7.2)   $a_{-1}=-J_0, \ b_{-1}=I_0,$ and $c_{-1}=K_0$;

\hspace{0.42cm}  (7.3)   $a_{-1}=-J_0+const.,$ and $
b_{-1}=b_{-1}(J_0),c_{-1}=c_{-1}(J_0)$ satisfy $
     \frac{d}{dJ_0}(b_{-1}c_{-1})=2a_{-1}$;

\hspace{0.42cm}  (7.4)   $a_{-1}=-J_0+const., \ b_{-1}=I_0,$ and $
  c_{-1}=c_{-1}(J_0)$;

\hspace{0.42cm}  (7.5)   $a_{-1}=-J_0+const., \ c_{-1}=-K_0,$ and
$ b_{-1}=b_{-1}(J_0)$.}

8.   $a_{-2}=a_{-1}=b_{-1}=c_{-1}=0$, $a_0,\ a_1=const.,\ b_0\ne 0,\ c_0\ne 0$,
    and $b_0, c_0$ satisfy the relation $\frac{d}{dJ_0}(b_0c_0)=-2a_1.$

9.   $a_{-2}=a_{-1}=b_{-1}=c_{-1}=0$, $c_0,\ a_1$, $a_0=const.,$ and $b_0\ne 0 $.

10.   $a_{-2}=b_{-1}=c_0=a_1=0$, $a_{-1},\ a_0=const., \
    c_{-1}=-K_0+h(J_0)$, and $b_0$ satisfies the relation
    $\frac{d}{dJ_0}(b_0h(J_0))=-2a_0$.\\
\vspace{0.2cm}
The above all functions $f, \ g, \ h, \ f_i, \ g_i \ (i=1,2)$ are in $C^{\infty}(R)$.}

    {\bf Proof}\quad    Through some calculations we have {\small
\begin{eqnarray*}
\{A(\lambda),A(\mu)\}&=&2\frac{\partial a_{-2}}{\partial I_1}<\Lambda p,p>
                         (\frac{\lambda}{\mu^2}B(\lambda)-\frac{\mu}{\lambda^2}B(\mu))\\
                       & &+2\frac{\partial a_{-2}}{\partial I_1}<\Lambda q,q>
                         (\frac{\lambda}{\mu^2}C(\lambda)-\frac{\mu}{\lambda^2}C(\mu))\\
                       & & +
                        2\frac{\partial a_{-2}}{\partial I_1}
                         (\frac{1}{\lambda^2}-\frac{1}{\mu^2})(<\Lambda p,p>
                         (b_{-1}-I_0)-<\Lambda q,q>(c_{-1}+K_0))\\
                       & & +
                        2\frac{\partial a_{-2}}{\partial I_1}
                         (\frac{\mu}{\lambda^2}-\frac{\lambda}{\mu^2})(<\Lambda p,p>
                         b_0+<\Lambda q,q>c_0),\\
\{B(\lambda),B(\mu)\}&=&2\frac{\partial b_{-1}}{\partial J_0}
                         (\frac{1}{\mu}B(\lambda)-\frac{1}{\lambda}B(\mu))
                        +
                        2\frac{d b_0}{d J_0}
                         (B(\lambda)-B(\mu))\\
                       & & +2
                         (\frac{1}{\lambda}-\frac{1}{\mu})
                        (\frac{\partial b_{-1}}{\partial I_0}\frac{db_0}{dJ_0}I_0
                        +b_0\frac{\partial b_{-1}}{\partial J_0}-b_{-1}\frac{db_0}{dJ_0}),\\
\{C(\lambda),C(\mu)\}&=&2\frac{\partial c_{-1}}{\partial J_0}
                         (-\frac{1}{\mu}C(\lambda)+\frac{1}{\lambda}C(\mu))
                        +
                        2\frac{d c_0}{d J_0}
                         (-C(\lambda)+C(\mu))\\
                       & & +2
                         (-\frac{1}{\lambda}+\frac{1}{\mu})
                        (\frac{\partial c_{-1}}{\partial K_0}\frac{dc_0}{dJ_0}K_0
    +c_0\frac{\partial c_{-1}}{\partial J_0}-c_{-1}\frac{dc_0}{dJ_0}),\\
\{A(\lambda),B(\mu)\}&=&\frac{2}{\lambda-\mu}(-B(\mu)+B(\lambda))-\frac{2}{\lambda}
                           \frac{da_{-1}}{dJ_0}B(\mu)\\
                       & &   +\frac{2}{\mu}\frac{\partial b_{-1}}{\partial I_0}B(\lambda)
                       -\frac{2\mu}{\lambda^2}(\frac{\partial a_{-2}}{\partial J_1}
                         B(\mu)+2\frac{\partial a_{-2}}{\partial I_1}<\Lambda q,q>A(\mu))\\
                       & &-2<\Lambda q,q>(2\frac{\partial b_{-1}}{\partial I_0}\frac{\partial a_{-2}}{\partial I_1}J_1
                          +\frac{\partial b_{-1}}{\partial I_0}\frac{\partial a_{-2}}{\partial J_1}
                          +2a_{-2}\frac{\partial a_{-2}}{\partial I_1})\lambda^{-2}\mu^{-1}\\
                       & &+2(-\frac{\partial b_{-1}}{\partial I_0}\frac{da_{-1}}{dJ_0}I_0
                       -b_{-1}\frac{\partial b_{-1}}{\partial I_0}+b_{-1}\frac{da_{-1}}{dJ_0}+b_{-1})\lambda^{-1}\mu^{-1}\\
                       & &-2(2<\Lambda q,q>\frac{\partial a_{-2}}{\partial I_1}(J_0+a_{-1})
                          +\frac{\partial a_{-2}}{\partial J_1}(I_0-b_{-1}))\lambda^{-2}\\
                       & &  -2(-b_0\frac{\partial a_{-2}}{\partial J_1}+2a_0<\Lambda q,q>
                          \frac{\partial a_{-2}}{\partial I_1})\lambda^{-2}\mu\\
                       & &+2b_0\frac{da_{-1}}{dJ_0}\lambda^{-1}
                       -2b_0\frac{\partial b_{-1}}{\partial I_0}\mu^{-1}
                        -4a_1<\Lambda q,q>\frac{\partial a_{-2}}{\partial I_1}\lambda^{-2}\mu^{2},\\
\{A(\lambda),C(\mu)\}&=&\frac{2}{\lambda-\mu}(C(\mu)-C(\lambda))+\frac{2}{\lambda}
                           \frac{da_{-1}}{dJ_0}C(\mu)\\
                       & &   +\frac{2}{\mu}\frac{\partial c_{-1}}{\partial K_0}C(\lambda)
                       +\frac{2\mu}{\lambda^2}(\frac{\partial a_{-2}}{\partial J_1}
                         C(\mu)+2\frac{\partial a_{-2}}{\partial I_1}<\Lambda p,p>A(\mu))\\
                       & &+2<\Lambda p,p>(2\frac{\partial c_{-1}}{\partial K_0}\frac{\partial a_{-2}}{\partial I_1}J_1
                       +\frac{\partial c_{-1}}{\partial K_0}\frac{\partial a_{-2}}{\partial J_1}
                       -2a_{-2}\frac{\partial a_{-2}}{\partial I_1})\lambda^{-2}\mu^{-1}\\
                      & & +2(\frac{\partial c_{-1}}{\partial K_0}\frac{da_{-1}}{dJ_0}K_0
                       -c_{-1}\frac{\partial c_{-1}}{\partial K_0}-c_{-1}\frac{da_{-1}}{dJ_0}-c_{-1})\lambda^{-1}\mu^{-1}\\
                       & &-2(-2<\Lambda p,p>\frac{\partial a_{-2}}{\partial I_1}(J_0+a_{-1})
                          +\frac{\partial a_{-2}}{\partial J_1}(K_0+c_{-1}))\lambda^{-2}\\
                      & & -2(c_0\frac{\partial a_{-2}}{\partial J_1}+2a_0<\Lambda p,p>
                          \frac{\partial a_{-2}}{\partial I_1})\lambda^{-2}\mu\\
                       & &-2c_0\frac{da_{-1}}{dJ_0}\lambda^{-1}
                       -2c_0\frac{\partial c_{-1}}{\partial K_0}\mu^{-1}
                        -4a_1<\Lambda p,p>\frac{\partial a_{-2}}{\partial I_1}\lambda^{-2}\mu^{2},\\
\{B(\lambda),C(\mu)\}&=&\frac{4}{\lambda-\mu}(-A(\mu)+A(\lambda))+\frac{4}{\lambda}
                           \frac{\partial b_{-1}}{\partial I_0}A(\mu)\\
                       & &   -\frac{4}{\mu}\frac{\partial c_{-1}}{\partial K_0}A(\lambda)
                       +2(\frac{1}{\mu}\frac{\partial c_{-1}}{\partial J_0}+\frac{dc_0}{dJ_0})B(\lambda)
                        +2(\frac{1}{\lambda}\frac{\partial b_{-1}}{\partial J_0}+\frac{db_0}{dJ_0})C(\mu)\\
                       & & +4(-\frac{\partial b_{-1}}{\partial I_0}+1)a_{-2}\mu^{-2}\lambda^{-1}
                        +4(\frac{\partial c_{-1}}{\partial K_0}+1)a_{-2}\lambda^{-2}\mu^{-1}\\
                       & &+2(\frac{\partial c_{-1}}{\partial J_0}\frac{\partial b_{-1}}{\partial I_0}I_0
                        +2\frac{\partial b_{-1}}{\partial I_0}\frac{\partial c_{-1}}{\partial K_0}J_0
                        +\frac{\partial b_{-1}}{\partial J_0}\frac{\partial c_{-1}}{\partial K_0}K_0\\
                       & &-\frac{\partial b_{-1}}{\partial J_0}c_{-1}-\frac{\partial c_{-1}}{\partial J_0}b_{-1}
                        -2\frac{\partial b_{-1}}{\partial I_0}a_{-1}
                        +2\frac{\partial c_{-1}}{\partial K_0}a_{-1}+2a_{-1})\lambda^{-1}\mu^{-1}\\
                       & &+2(-\frac{\partial b_{-1}}{\partial J_0}c_{0}+\frac{\partial b_{-1}}{\partial I_0}\frac{dc_0}{dJ_0}I_{0}
                          -\frac{dc_0}{dJ_0}b_{-1}-2\frac{\partial b_{-1}}{\partial I_0}a_{0})\lambda^{-1}\\
                       & &   +2(-\frac{\partial c_{-1}}{\partial J_0}b_{0}+\frac{\partial c_{-1}}{\partial K_0}\frac{db_0}{dJ_0}K_{0}
                          -\frac{db_0}{dJ_0}c_{-1}+2\frac{\partial c_{-1}}{\partial K_0}a_{0})\mu^{-1}\\
                       & &-2(\frac{db_0}{dJ_0}c_0+\frac{dc_0}{dJ_0}b_0+2a_1)
                          +4\frac{\partial c_{-1}}{\partial K_0}a_1\mu^{-1}\lambda
                          -4\frac{\partial b_{-1}}{\partial I_0}a_1\lambda^{-1}\mu.
\end{eqnarray*}}

    According to {\bf Assumption (A)}, the terms that do not contain
$A(\lambda),\ A(\mu),\ B(\lambda),\\ B(\mu),
C(\lambda),\ C(\mu)$ in the above six equalities, are zero.
After discussing these terms, we can obtain every result
in Lemma 2.1. \hfill $\rule{2mm}{4mm}$

\section{Generalized r-matrix structure
and integrable Hamiltonian systems}
\setcounter{equation}{0}
     Let $L_{1}(\lambda)=L(\lambda)\otimes I$, $L_{2}(\mu)=I\otimes L(\mu)$.
In the following, we search for a general $4\times 4$ {\it r}-matrix
structure
$r_{12}(\lambda,\mu)$ such that the fundamental Poisson bracket \cite{FT}:
\begin{equation}
\{L(\lambda) \stackrel{\otimes}{,}  L(\mu)\}=[r_{12}(\lambda,\mu), L_{1}(\lambda)]-[r_{21}(\mu,\lambda), L_{2}(\mu)]
\end{equation}
holds, where $r_{21}(\lambda,\mu)=Pr_{12}(\lambda,\mu)P,\
P=\frac{1}{2}\sum_{i=0}^3\sigma_i \otimes \sigma_i$, and
$\sigma_i's$ are the standard Pauli matrices. For the given
Lax matrix (2.1) and the Poisson bracket (1.1),
we have the following Theorem.

   {\bf Theorem 3.1} \quad {\it Under {\bf Assumption (A)},
\be
r_{12}(\lambda,\mu)=\frac{2}{\mu-\lambda}P+S \label{r}
\ee
is an {\it r}-matrix structure satisfying (3.1), where {\scriptsize
\begin{eqnarray*}
S=\left(\begin{array}{cccc}
\frac{2\lambda}{\mu^2}\frac{\partial a_{-2}}{\partial J_1}+\frac{2}{\mu}\frac{da_{-1}}{d J_0}
  &\frac{2}{\mu}\frac{\partial b_{-1}}{\partial J_0} &
\frac{2\lambda}{\mu^2}<\Lambda q,q>\frac{\partial a_{-2}}{\partial I_1} & 0 \\
2\frac{dc_0}{dJ_0} & 0 &
\frac{2}{\mu}\frac{\partial c_{-1}}{\partial K_0} &
-\frac{2\lambda}{\mu^2}<\Lambda q,q>\frac{\partial a_{-2}}{\partial I_1} \\
-\frac{2\lambda}{\mu^2}<\Lambda p,p>\frac{\partial a_{-2}}{\partial I_1} &
-\frac{2}{\mu}\frac{\partial b_{-1}}{\partial I_0} & 0 & -2\frac{db_0}{dJ_0}\\
0 & \frac{2\lambda}{\mu^2}<\Lambda p,p>\frac{\partial a_{-2}}{\partial I_1} &
-\frac{2}{\mu}\frac{\partial c_{-1}}{\partial J_0} &
\frac{2\lambda}{\mu^2}\frac{\partial a_{-2}}{\partial J_1}+\frac{2}{\mu}\frac{da_{-1}}{d J_0}
\end{array}\right).
\end{eqnarray*}}
}

   {\bf Proof}\quad    Under {\bf Assumption (A)}, we have
\begin{eqnarray*}\{A(\lambda),A(\mu)\}&=&2\frac{\partial a_{-2}}{\partial I_1}<\Lambda p,p>
                         (\frac{\lambda}{\mu^2}B(\lambda)-\frac{\mu}{\lambda^2}B(\mu))\\
                        & &+2\frac{\partial a_{-2}}{\partial I_1}<\Lambda q,q>
                         (\frac{\lambda}{\mu^2}C(\lambda)-\frac{\mu}{\lambda^2}C(\mu)),\\
\{B(\lambda),B(\mu)\}&=&2\frac{\partial b_{-1}}{\partial J_0}
                         (\frac{1}{\mu}B(\lambda)-\frac{1}{\lambda}B(\mu))
                        +
                        2\frac{d b_0}{d J_0}
                         (B(\lambda)-B(\mu)),\\
\{C(\lambda),C(\mu)\}&=&2\frac{\partial c_{-1}}{\partial J_0}
                         (-\frac{1}{\mu}C(\lambda)+\frac{1}{\lambda}C(\mu))
                        +
                        2\frac{d c_0}{d J_0}
                         (-C(\lambda)+C(\mu)),\\
\{A(\lambda),B(\mu)\}&=&\frac{2}{\lambda-\mu}(-B(\mu)+B(\lambda))-\frac{2}{\lambda}
                           \frac{da_{-1}}{dJ_0}B(\mu)
                       +\frac{2}{\mu}\frac{\partial b_{-1}}{\partial I_0}B(\lambda)\\
                       & &-\frac{2\mu}{\lambda^2}(\frac{\partial a_{-2}}{\partial J_1}
                         B(\mu)-2\frac{\partial a_{-2}}{\partial I_1}<\Lambda q,q>A(\mu)),\\
\{A(\lambda),C(\mu)\}&=&\frac{2}{\lambda-\mu}(C(\mu)-C(\lambda))+\frac{2}{\lambda}
                           \frac{da_{-1}}{dJ_0}C(\mu)
                          +\frac{2}{\mu}\frac{\partial c_{-1}}{\partial K_0}C(\lambda)\\
                       & &+\frac{2\mu}{\lambda^2}(\frac{\partial a_{-2}}{\partial J_1}
                         C(\mu)+2\frac{\partial a_{-2}}{\partial I_1}<\Lambda p,p>A(\mu)),\\
\{B(\lambda),C(\mu)\}&=&\frac{4}{\lambda-\mu}(-A(\mu)+A(\lambda))+\frac{4}{\lambda}
                           \frac{\partial b_{-1}}{\partial I_0}A(\mu)
                          -\frac{4}{\mu}\frac{\partial c_{-1}}{\partial K_0}A(\lambda)\\
                       & &+2(\frac{1}{\mu}\frac{\partial c_{-1}}{\partial J_0}+\frac{dc_0}{dJ_0})B(\lambda)
                        +2(\frac{1}{\lambda}\frac{\partial b_{-1}}{\partial J_0}+\frac{db_0}{dJ_0})C(\mu).
\end{eqnarray*}
which completes the proof. \hfill $\rule{2mm}{4mm}$

In general, Eq. (\ref{r}) is a dynamical $r$-matrix structure, i.e. dependent
on canonical variables $p_i,q_i$ \cite{BV}.


  Now, we turn to consider the determinant of $L(\lambda)$
\begin{eqnarray}
-\det L(\lambda)&=&\frac{1}{2}TrL^{2}(\lambda)
                    =A^2(\lambda)+B(\lambda)C(\lambda) \nonumber\\
                &=&\sum_{i=-4}^{2}H_i\lambda^{i}+
                \sum_{j=1}^{N}\frac{E_{j}}{\lambda-\lambda_{j}}
\end{eqnarray}
where{\small
\begin{eqnarray}
H_{-4}&=&a_{-2}^2,\\
H_{-3}&=&2a_{-2}a_{-1},\\
H_{-2}&=&a_{-1}^2+2a_{-2}a_0+b_{-1}c_{-1}-2a_{-2}<\Lambda^{-1}p,q>,\\
H_{-1}&=&2a_{-2}a_1+2a_{-1}a_0+b_{-1}c_{0}+b_{0}c_{-1}
         -2a_{-2}<\Lambda^{-2}p,q> \nonumber \\
      & &-2a_{-1}<\Lambda^{-1}p,q>-b_{-1}<\Lambda^{-1}p,p>+c_{-1}<\Lambda^{-1}q,q>,\\
H_{0}&=&a_{0}^2+2a_{-1}a_1+b_{0}c_{0}+2a_{1}<p,q>,\\
H_{1}&=&2a_{0}a_{1},\\
H_{2}&=&a_{1},\\
E_{j}&=&(2a_{-2}\lambda_{j}^{-2}+2a_{-1}\lambda_{j}^{-1}
          +2a_{0}+2a_{1}\lambda_{j})p_jq_j \nonumber \\
     & &+(b_{-1}\lambda_{j}^{-1}+b_{0})p_j^2
        -(c_{-1}\lambda_{j}^{-1}+c_{0})q_j^2-\Gamma_j,\\
     & &\Gamma_j=\sum_{k=1,k\neq j}^{N}\frac{(p_{j}q_{k}-
p_{k}q_{j})^{2}}{\lambda_{j}-\lambda_{k}},  \  j=1,2,...,N. \label{Ej}
\end{eqnarray}}
Let (3.3) be multiplied by a fixed multiplier $\lambda^k$ ($k\in Z$),
then it leads to{\small
\begin{eqnarray}
\frac{1}{2}\lambda^k\cdot TrL^{2}(\lambda)
                &=&\sum_{l=-4}^{2}H_l\lambda^{l+k}+\sum_{i=0}^{k-1}
                    F_i\lambda^{k-1-i}+
                    \sum_{j=1}^{N}\frac{\lambda^k_jE_{j}}
                    {\lambda-\lambda_{j}}\nonumber\\
                &=&\sum_{l=k-4}^{-1}H_{l-k}\lambda^{l}+\sum_{l=0}^{k-1}
                       (H_{l-k}+F_{k-1-l})\lambda^{l} \nonumber \\
                & &      +\sum_{l=k}^{k+2}H_{l-k}\lambda^{l}
                +\sum_{j=1}^{N}\frac{\lambda^k_jE_{j}}{\lambda-\lambda_{j}}
                \label{Tr}
\end{eqnarray}}
where
\begin{equation}
F_m=\sum_{j=1}^{N}\lambda^m_jE_j,\ m=0,1,2,\ldots
\end{equation}
which reads{\small
\begin{eqnarray}
F_{m}&=&2a_{-2}<\Lambda^{m-2}p, q>+2a_{-1}<\Lambda^{m-1}p,q>
          +2a_{0}<\Lambda^mp,q>+2a_{1}<\Lambda^{m+1}p,q> \nonumber\\
     & &b_{-1}<\lambda^{m-1}p,p>+b_{0}<\Lambda^mp,p>
        -c_{-1}<\Lambda^{m-1}q,q>-c_{0}<\Lambda^mq,q>\nonumber\\
     & &-\sum_{i+j=m-1}(<\Lambda^ip,p><\Lambda^jq,q>-<\Lambda^ip,q><\Lambda^jp,q>).
\end{eqnarray}}

Because there is an {\it r}-matrix structure
satisfying (3.1),
\begin{equation}
\{L^2(\lambda) \stackrel{\otimes}{,} L^{2}(\mu)\}=[\bar{r}_{12}(\lambda,\mu), L_{1}(\lambda)]-
[\bar{r}_{21}(\mu,\lambda), L_{2}(\mu)],
\end{equation}
where
\begin{equation}
\bar{r}_{ij}(\lambda,\mu)=\sum_{k=0}^{1}\sum_{l=0}^{1}L_{1}^{1-k}(\lambda)
L_{2}^{1-l}(\mu) \cdot r_{ij}(\lambda,\mu)\cdot L_{1}^{k}(\lambda)
L_{2}^{l}(\mu), \quad  ij=12, 21.
\end{equation}
Thus,
\begin{equation}
4\{TrL^{2}(\lambda), TrL^{2}(\mu)\}=Tr\{L^{2}(\lambda) \stackrel{\otimes}{,}
L^{2}(\mu)\}=0.
\end{equation}
So, by (\ref{Tr}) we immediately obtain

    {\bf Theorem 3.2} \quad {\it Under {\bf assumption (A)}, the following equalities
\begin{eqnarray}
& &\{E_{i}, E_{j}\}=0, \  \{H_l, E_{j}\}=0, \   \{F_m, E_{j}\}=0, \\
& & i, j=1,2,\ldots,N,\ l=-4,\ldots,2,\ m=0,1,2,\dots, \nonumber
\end{eqnarray}
hold. Hence, the Hamiltonian systems $(H_l)$ and $(F_m)$
\begin{eqnarray}
(H_{l})&:& \quad q_x=\frac{\partial H_{l}}{\partial p}, \   p_x=
-\frac{\partial H_{l}}{\partial q}, \  l=-4,\ldots,2, \\
(F_{m})&:& \quad q_{t_{m}}=\frac{\partial F_{m}}{\partial p}, \   p_{t_{m}}=
-\frac{\partial F_{m}}{\partial q}, \quad   m=0,1,2,\ldots,
\end{eqnarray}
are completely integrable in Liouville's sense.}

    {\bf Corollary 3.1} \quad {\it All composition functions $f(H_l,F_m)$, $f\in
C^{\infty}(R)$ are completely integrable Hamiltonians in Liouville's sense.}

\section{Reductions}
\setcounter{equation}{0}
    For the various cases of Lemma 2.1,
we give the corresponding reductions of {\it r}-matrix structure $r_{12}(\lambda,\mu)$
in this section.
The following numbers of title coincide with the ones in
Lemma 2.1, i.e. the corresponding conditions are coincidental.


Before giving our reductions, we'd like to
re-stress the two ``terminologies" used usually in the theory of
integrable systems in order to avoid some confusions:
one is ``{\it constrained system}", which means the finite dimensional
Hamiltonian system
or symplectic map in $R^{2N}$ under the {\it Bargmann-type} constraint;
the other ``{\it restricted system}", which means the finite
dimensional Hamiltonian
system or symplectic map on some symplectic submanifold in $R^{2N}$ under the
{\it Neumann-type} constraint. In the future we shall follow this principle.

1. { \it
\begin{equation}
r_{12}(\lambda,\mu)=\frac{2\lambda}{\mu(\mu-\lambda)}P+
  2\frac{\partial a_{-2}}{\partial J_1}\cdot
  \frac{\lambda}{\mu^2}S+2\frac{\partial a_{-2}}{\partial I_1}\cdot
  \frac{\lambda}{\mu^2}Q,
\end{equation}
{\footnotesize
\begin{eqnarray*}
S=\left(\begin{array}{cccc}
1 & 0 & 0 & 0 \\
0 & 0 & 0 & 0 \\
0 & 0 & 0 & 0 \\
0 & 0 & 0 & 1
\end{array}\right), \
Q=\left(\begin{array}{cccc}
0 & 0 & <\Lambda q,q> & 0 \\
0 & 0 & 0 & -<\Lambda q,q> \\
-<\Lambda p,p> & 0 & 0 & 0 \\
0 & <\Lambda p,p> & 0 & 0
\end{array}\right).
\end{eqnarray*}}
Particularly, with $f(a_{-2})=-1$, (4.1) exactly reads as the {\it r}-matrix of
the constrained WKI (c-WKI) system \cite{Q2}.
With $a_{-2}=const.\ne 0$, (3.2) reads
the {\it r}-matrix $r_{12}(\lambda,\mu)\\=\frac{2\lambda}{\mu(\mu-\lambda)}P$
of ellipsoid geodesic flow \cite{KH}, or reads as
$$r_{12}(\lambda,\mu)=\frac{2}{\mu-\lambda}P+\frac{2}{\mu}S,\quad
S=\left(\begin{array}{cccc}
       0 & f'_1 & 0 & 0 \\
0 & 0 & -1 & 0 \\
0 & -1 & 0 & 0 \\
0 & 0 & -g'_1 & 0
\end{array}\right), $$
which is a new $r$-matrix structure. 
For simplicity, below write `$\prime$'$=\frac{d\ }{dJ_0}$.

2.\begin{equation}
r_{12}(\lambda,\mu)=\frac{2}{\mu-\lambda}P.
\end{equation}
This is nothing but the {\it r}-matrix of the well-known
constrained AKNS (c-AKNS) system \cite{C1}.

3.\begin{equation}
r_{12}(\lambda,\mu)=\frac{2}{\mu-\lambda}P+\frac{2}{\mu}S,\quad
S=\left(\begin{array}{cccc}
a_{-1}' & 0 & 0 & 0 \\
0 & 0 & -1 & 0 \\
0 & 0 & 0 & 0 \\
0 & 0 & 0 & a_{-1}'
\end{array}\right), \ a_{-1}'\ne 0.
\end{equation}
In particular, with $a_{-1}=-J_0$ Eq. (4.3) reads as the {\it r}-matrix of
the constrained LZ (c-LZ) system \cite{CG1}.

4.\begin{equation}
r_{12}=\frac{2}{\mu-\lambda}P+c_0'S,\quad S=\left(\begin{array}{cccc}
0 & 0 & 0 & 0 \\
1 & 0 & 0 & 0 \\
0 & 0 & 0 & 0 \\
0 & 0 & -1 & 0
\end{array}\right).
\end{equation}
With $c_0=-2\sqrt{J_0}$, Eq. (4.4) reads as
the {\it r}-matrix of the
constrained Hu (c-H) system \cite{CG1}.

5.\begin{equation}
r_{12}(\lambda,\mu)=\frac{2}{\mu-\lambda}P+S,\quad S=\left(\begin{array}{cccc}
   0 & \frac{1}{\mu}g' & 0 & 0 \\
c_0' & 0 & 0 & 0 \\
0 & -\frac{2}{\mu} & 0 & -\frac{1}{\mu}g' \\
0 & 0 & c_0' & 0
\end{array}\right).
\end{equation}
With $b_{-1}=I_0,\ c_0=const.$, Eq. (4.5) reads as
the {\it r}-matrix of the constrained Qiao (c-Q) system \cite{Q3}.

6.\begin{equation}
r_{12}(\lambda,\mu)=\frac{2}{\mu-\lambda}P+\frac{2}{\mu}S, \quad S=\left(\begin{array}{cccc}
1 & 0 & 0 & 0 \\
0 & 0 & 0 & 0 \\
0 & 1 & 0 & 0 \\
0 & 0 & 0 & 1
\end{array}\right).
\end{equation}
This is a new $r$-matrix.

7.

  (7.1)\begin{equation}
r_{12}(\lambda,\mu)=\frac{2}{\mu-\lambda}P+\frac{2}{\mu}S, \quad S=\left(\begin{array}{cccc}
0 & g'_2 & 0 & 0 \\
f'_2 & 0 & -1 & 0 \\
0 & -1 & 0 & 0 \\
0 & 0 & 0 & 0
\end{array}\right).
\end{equation}
This is also a new $r$-matrix.

  (7.2)\begin{equation}
r_{12}(\lambda,\mu)=\frac{2\lambda}{\mu(\mu-\lambda)}P.
\end{equation}
This is the {\it r}-matrix of the constrained Heisenberg spin chain (c-HSC)
system \cite{Q4}.

  (7.3)\begin{equation}
r_{12}(\lambda,\mu)=\frac{2}{\mu-\lambda}P-\frac{2}{\mu}S, \quad S=\left(\begin{array}{cccc}
1 & 0 & 0 & 0 \\
0 & 0 & 0 & 0 \\
0 & 0 & 0 & b'_{-1}\\
0 & 0 & c'_{-1} & 1
\end{array}\right).
\end{equation}
With $b'_{-1}=-1,\
c'_{-1}=1$, Eq. (4.9) becomes the {\it r}-matrix of the constrained
Levi (c-L) system \cite{Q7}.

  (7.4)  \begin{equation}
r_{12}(\lambda,\mu)=\frac{2}{\mu-\lambda}P-\frac{2}{\mu}S, \quad S=\left(\begin{array}{cccc}
1 & 0 & 0 & 0 \\
0 & 0 & 0 & 0 \\
0 & 1 & 0 & 0 \\
0 & 0 & c'_{-1} & 1
\end{array}\right).
\end{equation}
This is a new $r$-matrix.

  (7.5)  \begin{equation}
r_{12}(\lambda,\mu)=\frac{2}{\mu-\lambda}P-\frac{2}{\mu}S, \quad S=\left(\begin{array}{cccc}
1 & 0 & 0 & 0 \\
0 & 0 & 1 & 0 \\
0 & 0 & 0 & b'_{-1}\\
0 & 0 & 0 & 1
\end{array}\right).
\end{equation}
This is also a new $r$-matrix.

8. \begin{equation}
r_{12}(\lambda,\mu)=\frac{2}{\mu-\lambda}P+S, \quad S=\left(\begin{array}{cccc}
0 & 2b'_0 & 0 & 0 \\
2c'_0 & 0 & 0 & 0 \\
0 & 0 & 0 & 0 \\
0 & 0 & 0 & 0
\end{array}\right).
\end{equation}
With $b_0=c_0=\sqrt{J_0},\ a_1=-\frac{1}{2}$, Eq. (4.12) reads as
the {\it r}-matrix of the constrained Tu (c-T)
system \cite{CG1}.

9.  \begin{equation}
r_{12}(\lambda,\mu)=\frac{2}{\mu-\lambda}P+S, \quad S=b'_0\left(\begin{array}{cccc}
0 & 1 & 0 & 0 \\
0 & 0 & 0 & 0 \\
0 & 0 & 0 & -1 \\
0 & 0 & 0 & 0
\end{array}\right).
\end{equation}
With $b_0=-J_0$, Eq. (4.13) reads as the common {\it r}-matrix
of the
constrained Toda (c-Toda) system (a discrete system) and
the constrained CKdV (c-CKdV) system (a continuous system),
which can be seen in subsection 5.1.

10.   \begin{equation}
r_{12}(\lambda,\mu)=\frac{2}{\mu-\lambda}P+S, \quad
S=\left(\begin{array}{cccc}
0 & b'_0 & 0 & 0 \\
\frac{1}{\mu}h' & 0 & -\frac{2}{\mu} & 0 \\
0 & 0 & 0 & -b'_0 \\
0 & 0 & -\frac{1}{\mu}h' & 0
\end{array}\right).
\end{equation}
With $h=const.,\ b_0=0$, Eq. (4.14) reads as
the {\it r}-matrix  of the constrained MKdV (c-MKdV)
system \cite{Q1}.}

    {\bf Proof}\quad  For simplicity, we only present the proof
    in cases 1 and 10, other cases are similar.

{\it Case 1}: \ \  With $a_{-2}\not =const.$,
the matrix $S$ becomes {\scriptsize
$$
 S=\left(\begin{array}{cccc}
\frac{2\lambda}{\mu^2}\frac{\partial a_{-2}}{\partial J_1}-\frac{2}{\mu}
  & 0  &
\frac{2\lambda}{\mu^2}<\Lambda q,q>\frac{\partial a_{-2}}{\partial I_1} & 0 \\
 0 & 0 &
-\frac{2}{\mu} &
-\frac{2\lambda}{\mu^2}<\Lambda q,q>\frac{\partial a_{-2}}{\partial I_1} \\
-\frac{2\lambda}{\mu^2}<\Lambda p,p>\frac{\partial a_{-2}}{\partial I_1} &
-\frac{2}{\mu} & 0 & 0\\
0 & \frac{2\lambda}{\mu^2}<\Lambda p,p>\frac{\partial a_{-2}}{\partial I_1} &
0 & \frac{2\lambda}{\mu^2}\frac{\partial a_{-2}}{\partial J_1}-\frac{2}{\mu}
\end{array}\right).$$}
Substituting $S$ into (3.2) and sorting it, we can
obtain (4.1), where $a_{-2}$ satisfies the relation $I_1=(J_1+a_{-2})^2+f(a_{-2})$,
for any $f(a_{-2})\in C^{\infty}(R)$. Particularly,
choosing $f(a_{-2})=-1$ yields $a_{-2}=\sqrt{1+<\Lambda p,p><\Lambda q,q>}
-<\Lambda p,q>.$ Thus Eq.
(4.1) reads
$$ r_{12}(\lambda,\mu)=\frac{2\lambda}{\mu(\mu-\lambda)}P-
  \frac{2\lambda}{\mu^2}S+\frac{\lambda}{\mu^2}\frac{1}{\sqrt{1+<\Lambda p,p><\Lambda q,q>}}Q,
\hspace{1cm}(r-WKI)  $$
while the corresponding Lax matrix $L(\lambda)$ becomes
$$  L(\lambda)=\left(\begin{array}{lr}
l_{11} & <q,q>\lambda^{-1} \\
-<p,p>\lambda^{-1} & -l_{11}
\end{array}\right)+\sum_{j=1}^{N}\frac{1}{\lambda-\lambda_{j}}\left(\begin{array}{lr}
p_{j}q_{j} & -q_{j}^{2} \\
p_{j}^{2} & -p_{j}q_{j}
\end{array}\right)$$
where
$$l_{11}=(\sqrt{1+<\Lambda p,p><\Lambda q,q>}-<\Lambda p,q>)\lambda^{-2}-<p,q>\lambda^{-1}.$$

Set an auxiliary matrix $M_1$ as follows
$$M_1=M_1(\lambda)=
\left(\begin{array}{cc}
-\lambda & \frac{<\Lambda q,q>}{\sqrt{1+<\Lambda p,p><\Lambda q,q>}}\lambda\\
-\frac{<\Lambda p,p>}{\sqrt{1+<\Lambda p,p><\Lambda q,q>}}\lambda & \lambda
\end{array}\right),$$
then the Lax equation
$$L_x=[M_1,L]$$
is equivalent to the following
finite dimensional Hamilton system $(\sqrt{H_{-4}})$:
\begin{equation}
(\sqrt{H_{-4}}):\qquad \left\{\begin{array}{l}
q_{x}=-\Lambda q+\frac{<\Lambda q,q>}{\sqrt{1+<\Lambda p,p><\Lambda q,q>}}\Lambda p=\frac{\partial \sqrt{H_{-4}}}{\partial p},  \\
p_{x}=\Lambda p-\frac{<\Lambda p,p>}{\sqrt{1+<\Lambda p,p><\Lambda q,q>}}\Lambda q=-\frac{\partial \sqrt{H_{-4}}}{\partial q},
\end{array}\right. \label{H4}
\end{equation}
with
$$\sqrt{H_{-4}}=a_{-2}=-<\Lambda p,q>+\sqrt{1+<\Lambda p,p><\Lambda q,q>},$$
which is obviously integrable by Theorem 3.2.

Let
\begin{equation}
u=\frac{<\Lambda q,q>}{\sqrt{1+<\Lambda p,p><\Lambda q,q>}},\  v=-\frac{<\Lambda p,p>}{\sqrt{1+<\Lambda p,p><\Lambda q,q>}},
\label{uv}
\end{equation}
then $(\sqrt{H_{-4}})$ is nothing but the WKI spectral problem \cite{WKI}
$$y_x=
\left(\begin{array}{cc}
-\lambda & \lambda u\\
\lambda v & \lambda
\end{array}\right)y $$
with the above two constraints (\ref{uv}), $\lambda=\lambda_j,$ and $y=(q_j,p_j)^T,
\ j=1,\ldots,N$.
That means that $(r-WKI)$ is the {\it r}-matrix of
the integrable constrained WKI (c-WKI) system (\ref{H4}).

    Other subcases in case 1 can be similarly proven.

    The proof of Case 10 can be found in ref. \cite{Q1}.
    In this case, the corresponding constrained system
    is reduced to the well-known MKdV spectral problem \cite{W}.
    \hfill $\rule{2mm}{4mm}$

    {\bf Remark 4.1} \quad From the above formulas (4.1)-(4.14), the
    {\it r}-matrices of cases 2, 6, and (7.2) are non-dynamical.
But in fact, for other cases we can also obtain
non-dynamical {\it r}-matrices if
choosing some special functions, for instance, in Eq. (4.3) setting $a_{-1}$
such that $a_{-1}'=const.$ leads to a non-dynamical one. Of course, we can
also get dynamical {\it r}-matrix,
for instance, in Eq. (4.4) choosing $c_0=-2\sqrt{J_0}$ yields a dynamical one.

    {\bf Remark 4.2} \quad  Equalities (4.1)-(4.14) cover most
    {\it r}-matrices of $2\times 2$ constrained systems.
 But among them there are also some new
    $r$-matrices and finite dimensional integrable systems
    like cases 6, 7.1, 7.4, and 7.5 (also see section 7).
    Their Lax matrices are altogether unified in Eq. (2.1).
    So, quite a large number
    of finite dimensional integrable systems are classified or
    reduced from the viewpoint of Lax matrix and $r$-matrix structure.

\section{Different systems sharing
the same  r-matrices}
\setcounter{equation}{0}

   In the above $r$-matrices, we
find some pairs of different integrable systems
sharing the common
$r$-matrices. Now, we present these results as follows.

\subsection{The constrained Toda and CKdV flows}
Let us consider
the following $2\times2$ traceless Lax matrix \cite{Qiao2}
(corresponding to case 9 in section 4)
\begin{eqnarray}
 L^{TC}&=&L^{TC}(\la,p,q)=\ju{-\frac{1}{2}\la}{<p,q>}{-1}{\frac{1}{2}\la}+
 L_0 \nonumber \\
 &\equiv&\ju{A_{TC}(\la)}{B_{TC}(\la)}{C_{TC}(\la)}{-A_{TC}(\la)}
  \label{CTLax}
 \end{eqnarray}
where
\be L_0=L_0(\la,p,q)=
 \sum_{j=1}^N\frac{1}{\la-\la_j}\ju{-p_jq_j}{p_j^2}{-q_j^2}{p_jq_j}. \label{l0}\ee
The determinant of Eq. (\ref{CTLax}) leads to
 \begin{eqnarray}
 \frac{1}{2}\la Tr(L^{TC})^2(\la)&=&-\frac{1}{2}\la^3+<p,q>\la+2H_C
 +\sum_{j=1}^N\frac{E^{TC}_j}{\la-\la_j},\\
E^{TC}_{j}&=&
\lambda_{j}p_jq_j-p_j^2-<p,q>q_j^2-\Gamma_j,  \  j=1,2,...,N, \label{Ej1}
\end{eqnarray}
where $\Gamma_j$ is defined by (\ref{Ej}) and the Hamiltonian function $H_C$ is
\be
H_C=-\frac{1}{2}<p,p>+\frac{1}{2}<\La q,p>-\frac{1}{2}<q,q><p,q>.
\ee
Viewing the variables $q$ and $p$ as the functions of continuous
variables $x$, then we have the following Hamiltonian
canonical equation $(H_C)$:
\be \left\{\begin{array}{l}
 p_x=-\frac{\pa H_C}{\pa q}=-\frac{1}{2}\La p+\frac{1}{2}<q,q>p+<p,q>q,\\
 q_x=\frac{\pa H_C}{\pa p}=-p+\frac{1}{2}\La q-\frac{1}{2}<q,q>q,
  \end{array}\right.\label{ch}\ee
which is nothing but
 the coupled KdV (CKdV) spectral problem \cite{LSR}
 \be \psi_x=\ju{-\frac{1}{2}\la+\frac{1}{2}u}{v}{-1}{\frac{1}{2}\la-
 \frac{1}{2}u}\psi
\label{csp} \ee
 with the two constraints (Bargmann-type)
\be u=<q,q>,\ \ v=<p,q>,\label{uv1}\ee
$\la=\la_j $ and $\psi=(p_j,q_j)^T.$ So,
$(H_C)$ coincides with the constrained CKdV (c-CKdV) flow.

Let us consider endowing with an
auxiliary $2\times2$ matrix $M_{T}$ as follows
\be
M_{T}=\ju{0}{g}{-\frac{1}{g}}{\frac{\la-<q,q>}{g}},\ \
g^2=<\La q,q>-<p,q>-<q,q>^2.\label{mt}\ee
Then, we have the following theorem.

{\bf Theorem 5.1.2}\quad {\it The discrete Lax equation
\be (L^{TC})'M_{T}=M_TL^{TC},\ \ (L^{TC})'=L^{TC}(\la, p', q') \label{dl}
\ee
is equivalent to a finite-dimensional symplectic map
$H_T: \ R^{2N}\longrightarrow R^{2N}, (p,q)\longmapsto (p',q')$,
which is called the constrained Toda (c-Toda) flow}:
\begin{equation}
 \left\{\begin{array}{l}
  p'=gq,\\
  q'=\frac{\Lambda q-p-<q,q>q}{g}.
 \end{array}\right. \label{ctd}
\end{equation}

    {\bf  Proof} \quad Directly calculate, and
    readily show $(\ref{dl})\Longleftrightarrow
    (\ref{ctd})$ and $(H_T)^{*}(dp\wedge dq)=dp\wedge dq$.
    \hfill $\rule{2mm}{4mm}$

    When we understand the above two matrices $(L^{TC})'$ and $M_T$ in the
following sense: $(L^{TC})' \longrightarrow L^{TC}_{n+1},\
M_{T}\longrightarrow M_{Tn}$ (i.e., $q\longrightarrow q_{n},
p\longrightarrow p_{n},$ here $n$ is the discrete variable),
and set
    \begin{equation}
 \left\{\begin{array}{l}
 u_n=\pm(<\Lambda q_n,q_n>-<p_n,q_n>-<q_n,q_n>^2)^{\frac{1}{2}},\\
 v_n=<q_n,q_n>,
 \end{array}\right. \label{uvn1}
 \end{equation}
then the constrained Toda flow (\ref{ctd}) is none other than
the well-known Toda spectral problem
\begin{equation}
L\psi_n\equiv (E^{-1}u_n+v_n+u_nE)\psi_n=\lambda \psi_n, \quad Ef_n=f_{n+1},\quad  E^{-1}f_n=f_{n-1}
\label{sp1}
\end{equation}
with the above constraint (\ref{uvn1}), $\lambda=\lambda_j$ and $\psi_n=q_{n,j}$.
Theorem 5.2 shows that the constrained Toda flow
$(H_T)$ has the discrete Lax representation (\ref{dl}).
Eq. (\ref{uvn1}) is a kind of discrete Bargmann constraint \cite{RCW}
of the Toda spectral problem (\ref{sp1}).

The Hamiltonian systems $(H_T)$ and
$(H_C)$ share the common Lax matrix (\ref{CTLax}). Thus, they
have the following same $r$-matrix:
\be
r_{12}(\lambda,\mu)=\frac{2}{\mu-\lambda}P-S, \ \ S=\left(\begin{array}{cccc}
0 & 1 & 0 & 0 \\
0 & 0 & 0 & 0 \\
0 & 0 & 0 & -1 \\
0 & 0 & 0 & 0
\end{array}\right)\label{rtc}
\ee
which is proven to satisfy the classical Yang-Baxter equation (YBE)
\be
[r_{ij},r_{ik}]+[r_{ij},r_{jk}]+[r_{kj},r_{ik}]=0,\ i,j,k=1,2,3.
\ee

\subsection{The restricted Toda and CKdV flows}
Let us now consider the case on a symplectic manifold. We
restrict the Toda and CKdV flows
on the following symplectic submanifold $M$ in $R^{2N}$
 \be M=\{(q,p)\in R^{2N}|F\equiv <q,q>-1=0, G\equiv <q,p>-\frac{1}{2}=0\}.\ee
Let us first introduce the Dirac bracket
 \be \{f,g\}_D=\{f,g\}+\frac{1}{2}(\{f,F\}\{G,g\}-\{f,G\}\{F,g\})
 \label{Dirac}\ee
which is easily proven to be a Poisson bracket on $M$.

According to the thought of ref. \cite{Qiao2},
the following Lax matrix
\be
 L^{TC}_R=L^{TC}_R(\la,p,q)=\ju{-\frac{1}{2}}{0}{0}{\frac{1}{2}}+L_0\equiv
 \ju{A_R(\la)}{B_R(\la)}{C_R(\la)}{-A_R(\la)}
      \label{RLax}\ee
yields
\be -\la^2\det L^{TC}_R=\frac{1}{2}\la^2Tr (L^{TC}_R)^2
=\frac{1}{4}\la^2+<p,q>\la+2H^{C}_R+\sum_{j=1}^N\frac{\la_j^2E^{TC}_{R,j}}{\la-\la_j}
\label{det}\ee
where
\begin{eqnarray}
 H^C_R&=&\frac{1}{2}<\La p,q>-\frac{1}{2}<q,q><p,p>+\frac{1}{2}<p,q>^2, \label{h}\\
 E^{TC}_{R,j}&=&E^{TC}_{R,j}(p,q)=p_{j}q_{j}-\Gamma_j,\ i=1,\ldots,N,\label{e}
\end{eqnarray}
and $L_0$ is defined by (5.2).

An important observation is: if we consider the {\bf Hamiltonian canonical equation produced by (\ref{h})
in ${\bf R^{2N}}$, then this equation is exactly the well-known constrained
AKNS flow}, which will be discussed in the next subsection.
Now, we first consider the {\bf Hamiltonian canonical equation
restricted on ${\bf M}$}:
\be (H^C_R):\ \  q_x=\{q,H^C_R\}_{D}, \ p_x=\{p,H^C_R\}_D, \label{hm} \ee
which reads as the following finite
dimensional system:
\begin{eqnarray}
 \left\{\begin{array}{l}
 p_x=-\frac{1}{2}\La p+\frac{1}{2}(<\La q,q>-1)p+<p,p>q,\\
 q_x=-p+\frac{1}{2}\La q-\frac{1}{2}(<\La q,q>-1)q,\\
 <q,q>=1,<q,p>=\frac{1}{2}.
 \end{array}\right. \label{rh} \end{eqnarray}
This is actually the CKdV spectral problem (\ref{csp})
 with the two constraints (Neumann-type) \cite{CG2}
 \be u=<\La q,q>-1, \ \ v=<p,p> \label{uv2} \ee
and $\la =\la_j,\ \psi=(p_j,q_j),\ j=1,2,...,N$.
So, the {\it
finite dimensional system (\ref{rh}) coincides with
the restricted CKdV (r-CKdV) flow.}

Let us return to the Lax matrix (\ref{RLax}). After endowing
with an auxiliary matrix $M_{T,R}$ as follows
\be
M_{T,R}=\ju{0}{a}{-\frac{1}{a}}{\frac{\la-b}{a}},\label{mn}\ee
$$a^2=<\Lambda q-p,\Lambda q-p>+<\Lambda q,q>-<\Lambda q,q>^2,$$
$$ b=<\La q,q>-1,$$
then we have the following theorem.

{\bf Theorem 5.2.1}\quad {\it
The discrete Lax equation
\be (L^{TC}_R)'M_{T,R}=M_{T,R}L^{TC}_R,\ \  (L^{TC}_R)'=
L^{TC}_R(\la,p',q') \label{dl1}
\ee
is equivalent to a discrete Neumann type of finite dimensional
symplectic map ${\cal H_T}: (p, q)^T
\rightarrow (p',q')^T$
\be \left\{\begin{array}{l}
  p'=aq,\\
  q'=a^{-1}(\La q-p-bq),\\
  <q,q>=1,\ <q,p>=\frac{1}{2},
 \end{array}\right.\label{de}\ee
which is called the restricted Toda (r-Toda) flow.}

{\bf Remark 5.2.1} \ \ If we understand the above two matrices
$(L^{TC}_R)'$ and
$M_{T,R}$ in
the following sense: $(L^{TC}_R)'\rightarrow (L^{TC}_R)_{n+1},\
 M_{T,R}\rightarrow (M_{T,R})_n$ (i.e.,
$q\rightarrow q_{n}, p\rightarrow p_{n}, a\rightarrow a_n, b\rightarrow b_n$,
here $n$ is the discrete variable), then the restricted Toda flow (\ref{de})
on the symplectic submanifold $M=\{(q,p)\in R^{2N}|<q,q>=1,
<q,p>=\frac{1}{2}\}$
is nothing but the discrete Neumann system studied by Ragnisco \cite{R}.

Let $L^{TC}_{R1}=L^{TC}_R(\la,p,q)\otimes I$ and $L^{TC}_{R2}=
I\otimes L^{TC}_R(\mu,p,q)$.
Then, under the Dirac bracket (\ref{Dirac})
we obtain the following theorem.

{\bf Theorem 5.2.2} \quad{\it The Lax matrix $L^{TC}_R(\la,p,q)$ defined
by (\ref{RLax}) satisfies the following
fundamental Dirac-Poisson bracket
\be \{L^{TC}_{R}(\la)\stackrel{\otimes}{,} L^{TC}_{R}(\mu)\}_D
=[r_{12}(\la,\mu), L^{TC}_{R1}(\la)]-[r_{21}(\mu,\la),L^{TC}_{R2}(\mu)]
\label{mr}\ee
with a dynamical $r$-matrix
\be r_{12}(\la,\mu)=\frac{2}{\mu-\la}P-S_{12}(\la,\mu), \ \
r_{21}(\mu,\la)=Pr_{12}(\mu,\la)P \ee
where
{\scriptsize \be
S_{12}=(E_{11}-E_{22})\otimes E_{12}+E_{11}\otimes\ju{C_R(\mu)}{0}{0}{0}\\
+E_{12}\otimes\ju{0}{-B_R(\mu)}{C_R(\mu)}{0}+E_{22}\otimes\ju{0}{2A_R(\mu)}
{0}{C_R(\mu)}
\ee}
and $P= \frac{1}{2}(I+\sum_{j=1}^3\sigma_j\otimes\sigma_j)$ is the permutation
matrix, $\sigma_j$ $(j=1,2,3)$ are the Pauli matrices.}

This Theorem ensures that (\ref{e}) satisfies
 \be \{E^{TC}_{R,i},E^{TC}_{R,j}\}_D=0, \ i,j,=1,\ldots,N. \label{eei}\ee
For the r-Toda flow (\ref{de}),
we have  $E^{TC}_{R,i}(p',q')=E^{TC}_{R,i}(p,q)$ as well as
$\sum_{i=1}^n E^{TC}_{R,i}=<p,q>=\frac{1}{2}$ from the discrete Lax
equation (\ref{dl1}).
Thus, in the set $\left\{E^{TC}_{R,j}\right\}_{j=1}^N$,
only $E^{TC}_{R,1},\ E^{TC}_{R,2},\ldots,\ E^{TC}_{R,N-1}$
are independent on $M$. Therefore we obtain the following Theorem.

 {\bf Theorem 5.2.3} \quad {\it
  The restricted Toda flow ${\cal H_T}$ is completely
integrable, and its independent and invariant (N-1)-involutive system
is $\{E^{TC}_{R,i}\}_{i=1}^{N-1}$.}

 For the  restricted CKdV flow on $M$, we have
 \be  H^C_R=\frac{1}{2}\sum_{j=1}^N\la_jE^{TC}_{R,j}\ee
which implies
\be \{H^C_R,E^{TC}_{R,j}\}_D=0,\quad j=1,2,\ldots,N.\ee
Thus, the following Theorem holds.

 {\bf Theorem 5.2.4} \quad {\it  The r-CKdV flow
 $(H^C_R)$  is completely integrable,
 and its independent (N-1)-involutive system is also
 $\{E^{TC}_{R,k}\}_{k=1}^{N-1}$.}

{\bf Remark 5.2.3} \ \  As shown in this and last subsection, the
r-Toda (i.e. Neumann-type) and the r-CKdV flows,
and the c-Toda (i.e. Bargmann-type) and the c-CKdV flows
respectively share the completely same Lax matrix, $r$-matrix and
involutive conserved integrals.
Thus, we say
that the finite dimensional integrable CKdV flow both restricted
and constrained
is the interpolating Hamiltonian flow of
invariant of the corresponding Toda integrable symplectic map.

\subsection{The constrained AKNS and Dirac (D) flows}
From now on we assume:
\be L_0=L_0(\la,p,q)=
 \sum_{j=1}^N\frac{1}{\la-\la_j}\ju{p_jq_j}{-q_j^2}{p_j^2}{-p_jq_j}.
 \label{l01}\ee

Let us again consider Eq. (5.18), and rewrite
it as the following version:
\begin{eqnarray}
L^{AKNS}&=&L^{AKNS}(\lambda,p,q)
=\left(\begin{array}{cc}
   1 & 0\\
   0 & -1
 \end{array}\right)
 +L_0, \label{LAKNS}
\end{eqnarray}
while we introduce
\begin{eqnarray}
L^{D}&=&L^{D}(\lambda,p,q)
=\left(\begin{array}{cc}
 0 & 1\\
 -1 & 0
 \end{array}\right)+L_0. \label{LD}
\end{eqnarray}
Then we have
 {\small \be
\frac{1}{2}\la^2Tr (L^{AKNS})^2(\la)
=\la^2+2\la<p,q>+<p,q>^2+2H_{AKNS}+
\sum_{j=1}^N\frac{\la_j^2E_j^{AKNS}}{\la-\la_j}, \label{detAKNS}\ee }
{\small \be \frac{1}{2}\la^2Tr (L^{D})^2(\la)
=\la^2+\la(<q,q>+<p,p>)-\frac{1}{4}(<p,p>+<q,q>)^2+2H_{D}+
\sum_{j=1}^N\frac{\la_j^2E^D_{j}}{\la-\la_j},
\label{detD}\ee}
where
\begin{eqnarray}
 H_{AKNS}&=&<\La p,q>-\frac{1}{2}<q,q><p,p>, \label{hAKNS}\\
 H_{D}&=&\frac{1}{2}(<\La q,q>+<\La p,p>)+\frac{1}{2}
        (<p,q>^2-<q,q><p,p>) \nonumber\\
      & & +\frac{1}{8}(<p,p>+<q,q>)^2, \label{hD}\\
 E_j^{AKNS}&=&2p_{j}q_{j}-\Gamma_j,  \ j=1,\ldots,N, \label{Ej3}\\
 E_j^{D}&=&p_{j}^2+q_{j}^2-\Gamma_j, \ j=1,\ldots,N.\label{Ej4}
\end{eqnarray}
Thus, $H_{AKNS}$ and $H_D$ generate the following two Hamiltonian systems
{\small
\begin{eqnarray}
(H_{AKNS})&:& \  \left\{\begin{array}{l}
q_{x}=\frac{\partial H_{AKNS}}{\partial p}=-<q, q>p+\Lambda q,  \\
p_{x}=-\frac{\partial H_{AKNS}}{\partial q}=<p, p>q-\Lambda p;
\end{array}
\right.\\
(H_{D})&:& \  \left\{\begin{array}{l}
q_{x}=\frac{\partial H_{D}}{\partial p}=<p, q>q+\frac{1}{2}(<p,p>-<q,q>)p+\Lambda p,  \\
p_{x}=-\frac{\partial H_{D}}{\partial q}=-<p, q>p-\frac{1}{2}(<p,p>-<q,q>)q-\Lambda q.
\end{array}
\right.
\end{eqnarray}}

It can be easily seen that $(H_{AKNS})$ and $(H_D)$ are
changed to the
well-known Zakharov-Shabat-AKNS spectral problem \cite{ZS}
\begin{equation}
y_{x}=\left(\begin{array}{cc}
 \lambda & u\\
 v & -\lambda
 \end{array}\right)y \label{AKNS}
\end{equation}
and
the Dirac spectral problem \cite{LS}
\begin{equation}
y_{x}=\left(\begin{array}{cc}
-v & \lambda-u\\
-\lambda-u & v
 \end{array}\right)y \label{DS}
 \end{equation}
with the constraints
$u=-<q,q>, \  v=<p, p>$, $\lambda=\lambda_j, \ y=(q_j, p_j)^T$, and
the constraints $u=-\frac{1}{2}(<p,p>-<q,q>), \ v=-<p, q>,$
$\lambda=\lambda_j, \ y=(q_j, p_j)^T$, respectively.

Therefore $(H_{AKNS})$ and $(H_D)$ coincide with the constrained AKNS
(c-AKNS) system and the constrained Dirac (c-D) system, respectively.

Let $L^J_{1}(\lambda)=L^J(\lambda)\otimes I$, $L^J_{2}(\mu)=I\otimes L^J(\mu)
\ (J=AKNS, D)$.
Then we have the following theorem.

{\bf  Theorem 5.3.1} \quad {\it
The Lax matrices $L^J(\lambda)\ (J=AKNS, D)$ defined by Eq. (\ref{LAKNS})
and Eq. (\ref{LD})
satisfy the fundamental Poisson bracket
\begin{equation}
\{L^J(\lambda) \stackrel {\otimes}{,} L^J(\mu)\}=[r_{12}(\lambda,\mu),
 L^J_{1}(\lambda)]-[r_{21}(\mu,\lambda), L^J_{2}(\mu)]. \label{fpb1}
\end{equation}
Here the $r$-matrices $r_{12}(\lambda,\mu)$, $r_{21}(\mu,\lambda)$
 are exactly given by the following standard $r$-matrix
\begin{equation}
r_{12}(\lambda,\mu)=\frac{2}{\mu-\lambda}P, \quad
r_{21}(\mu,\lambda)=Pr_{12}(\mu, \lambda)P,\label{rD}
\end{equation}
\begin{eqnarray}
P=\left(\begin{array}{cccc}
1 & 0 & 0 & 0 \\
0 & 0 & 1 & 0 \\
0 & 1 & 0 & 0 \\
0 & 0 & 0 & 1
\end{array}\right)=\frac{1}{2}(I+\sum_{i=1}^3\sigma_i\otimes\sigma_i).
\end{eqnarray}}

So, the c-AKNS and c-D flows share the same standard $r$-matrix
(\ref{rD}), which is obviously non-dynamical. However, {\bf
the two constrained flows, produced by
(\ref{AKNS})'s and (\ref{DS})'s extensive
spectral problems (\ref{GS}) and (\ref{QS}) (they are gauge equivalent),
have different $r$-matrices} (see section 6).

{\bf Remark 5.3.1}\quad
In fact, the $r$-matrix $r_{12}(\lambda,\mu)$
in the case of the c-AKNS and c-D flows
can be also chosen as
\begin{equation}
r_{12}(\lambda,\mu)=\frac{2}{\mu-\lambda}P+I\otimes \tilde{S},\ \tilde{S}=
\left(\begin{array}{cc}
a & b\\
c & d
\end{array}\right)
\end{equation}
where the elements $a, b, c, d$ can be arbitrary $C^{\infty}$-functions $a(\lambda,\mu,p,q),
b(\lambda,\mu,p,q),\\c(\lambda,\mu,p,q),d(\lambda,\mu,p,q)$
with respect to the spectral parametres $\lambda,\mu$
and the dynamical variables $p,q$.
This shows that for a given
Lax matrix, the associated $r$-matrix is not uniquely defined
(there are even infinitely many $r$-matrices possible).
Here we give the simplest case: $a=b=c=d=0$,
i. e. the standard $r$-matrix (\ref{rD}).

\subsection{\bf The constrained Harry-Dym (HD)
and Heisenberg spin chain (HSC) flows}
The constrained Harry-Dym system describes the geodesic flow on an
ellipsoid and shares the same $r$-matrix with the constrained Heisenberg
spin chain (HSC). To prove this, we consider the following Lax matrices:{\small
\begin{eqnarray}
L^{HD}&=&L^{HD}(\lambda,p,q)=\left(\begin{array}{cc}
 -<p,q>\lambda^{-1} & \lambda^{-2}+<q,q>\lambda^{-1}\\
 -<p,p>\lambda^{-1} & <p,q>\lambda^{-1}
 \end{array}\right)+L_0, \label{LaxHD}\\
 L^{HSC}&=&L^{HSC}(\lambda,p,q)=\left(\begin{array}{cc}
 -<p,q>\lambda^{-1} & <q,q>\lambda^{-1}\\
 -<p,p>\lambda^{-1} & <p,q>\lambda^{-1}
 \end{array}\right)+L_0. \label{LaxHSC}
\end{eqnarray}}
Here $L^{HSC}$ is included in the generalized Lax matrix (2.1),
but $L^{HD}$ is not.
We need two associated auxiliary matrices
\begin{eqnarray}
M_{HD}&=&
\left(\begin{array}{cc}
0 & 1\\
-\frac{<\Lambda p,p>}{<\Lambda^2q,q>}\lambda & 0
\end{array}\right),\\
M_{HSC}&=&
\left(\begin{array}{cc}
-i\lambda <\Lambda p,q> & i\lambda <\Lambda q,q>\\
-i\lambda <\Lambda p,p> & i\lambda <\Lambda p,q>
\end{array}\right), \quad i^2=-1.
\end{eqnarray}
{\bf Theorem 5.4.1} \quad {\it
The Lax representations
\begin{equation}
L_x^{HD}=[M_{HD}, L^{HD}],
\end{equation}
\begin{equation}
L_x^{HSC}=[M_{HSC}, L^{HSC}]
\end{equation}
respectively give the following fininte dimensional Hamiltonian flows:
\begin{eqnarray}
(H_{HD}): \  \left\{\begin{array}{l}
q_{x}=p=\frac{\partial H_{HD}}{\partial p}|_{TQ^{N-1}},  \\
p_{x}=-\frac{<\Lambda p,p>}{<\Lambda^2q,q>}\Lambda q=-\frac{\partial H_{HD}}
{\partial q}|_{TQ^{N-1}},\\
<\Lambda q,q>=1;
  \end{array}
  \right. \label{HD}
\end{eqnarray}
\begin{eqnarray}
(H_{HSC}): \  \left\{\begin{array}{l}
q_{x}=i<\Lambda q, q>\Lambda p-i<\Lambda p,q>\Lambda q=\frac{\partial H_{HSC}}{\partial p},\\
p_{x}=i<\Lambda p, q>\Lambda p-i<\Lambda p,p>\Lambda q=-\frac{\partial H_{HSC}}{\partial q},
\end{array}
\right.  \label{HSC}
\end{eqnarray}
with the Hamiltonian functions
 \begin{eqnarray}
H_{HD}&=&\frac{1}{2}<p,p>-\frac{<\Lambda p,p>}{2<\Lambda^2q,q>}(<\Lambda q,q>-1),\\
H_{HSC}&=&\frac{1}{2}i<\Lambda p,p><\Lambda q,q>-\frac{1}{2}i<\Lambda p,q>^2.
\end{eqnarray}
In Eq. (\ref{HD}) $TQ^{N-1}$ is a tangent bundle in $R^{2N}$:
\begin{eqnarray}
TQ^{N-1}=\{(p,q)\in R^{2N}|F\equiv <\Lambda q,q>-1=0, G\equiv <\Lambda p,q>=0\}.
\end{eqnarray}}
Obviously, Eq. (\ref{HD}) is equivalent to
\begin{equation}
q_{xx}+\frac{<\Lambda q_x,q_x>}{<\Lambda^2q,q>}\Lambda q=0, \
<\Lambda q,q>=1, \label{gf}
\end{equation}
which is nothing but the equation of the geodesic flow \cite{KH} on
the surface $<\Lambda q,q>=1$ in the space $R^N$ and also coincides
with the constrained
HD (c-HD) flow \cite{C4}.
In addition, Eq. (\ref{HSC}) becomes the
Heisenberg spin chain spectral problem \cite{T}
\begin{equation}
y_{x}=\left(\begin{array}{cc}
-i\lambda w & -i\lambda u\\
-i\lambda v & i\lambda w
 \end{array}\right)y,\quad i^2=-1,  \label{Heis}
 \end{equation}
with the constraints $
u=-<\Lambda q,q>, \ v=<\Lambda p,p>, \  w=-<\Lambda p, q>$,
$\lambda=\lambda_j, \ y=(q_j, p_j)^T$.
Thus, Eq. (\ref{HSC}) reads as the constrained
Heisenberg spin chain (c-HSC) flow \cite{Q4}.

Their Lax matrices (\ref{LaxHD}) and (\ref{LaxHSC})
share all elements except one,
namely
\[\left(\begin{array}{cc}
 0 & \lambda^{-2}\\
0 & 0
\end{array}\right)\,.\]
This element does not affect the calculations
concerning the fundamental Poisson bracket, one can
readily deduce that the c-HD flow and the
c-HSC flow possess the same non-dynamical $r$-matrix
\begin{equation}
r_{12}(\lambda,\mu)=\frac{2\lambda}{\mu(\mu-\lambda)}P, \
r_{21}(\mu,\lambda)=Pr_{12}(\mu, \lambda)P.\label{rHD}
\end{equation}

{\bf Remark 5.4.1}\quad
The $r$-matrix (\ref{rHD})
of the c-HD and c-HSC flows
can be also chosen as
\begin{equation}
r_{12}(\lambda,\mu)=\frac{2\lambda}{\mu(\mu-\lambda)}P+I\otimes \tilde{S}. \label{rHDS}
\end{equation}
Evidently, Eq. (\ref{rHD}) is the simplest case: $\tilde{S}=0$ of Eq.
(\ref{rHDS}).

\subsection{The constrained G and Q flows}
  In this subsection, we introduce the following Lax
matrices:{\small
\begin{eqnarray}
L^{G}&=&L^{G}(\lambda,p,q)=\left(\begin{array}{cc}
 (\frac{1}{2}+<p,q>)\lambda^{-1} & <q,q>\lambda^{-1}\\
 0 & -(\frac{1}{2}+<p,q>)\lambda^{-1}
 \end{array}\right)+L_0,\\
L^{Q}&=&L^{Q}(\lambda,p,q)=\left(\begin{array}{cc}
 -\lambda^{-1} & <q,q>\lambda^{-1}\\
 0 & \lambda^{-1}
 \end{array}\right)+L_0.
\end{eqnarray} }
If we set
\begin{equation}
M_{G}=
\left(\begin{array}{cc}
-\frac{1}{\alpha}\lambda & \frac{1}{\alpha}(<p,p>-<q,q>)-1\\
\frac{1}{\alpha}(<p,p>-<q,q>+1)\lambda & \frac{1}{\alpha}\lambda
\end{array}\right),
\end{equation}
\begin{equation}
M_{Q}=
\left(\begin{array}{cc}
\lambda+\frac{1}{2\beta^2}<\Lambda q,q><p,p> & \frac{1}{\beta}<\Lambda q,q>\\
-\frac{1}{\beta}<p,p>\lambda & -\lambda-\frac{1}{2\beta^2}<\Lambda q,q><p,p>
\end{array}\right),
\end{equation}
with
\be \alpha=\sqrt{(<p,p>-<\Lambda q,q>)^2-4<\Lambda q,p>}, \  \
\beta=1-<p,q>, \ee
then, by a lengthy and straightforward calculation we obtain the
following theorem.

{\bf Theorem 5.5.1} \quad {\it The following Lax representations
\begin{equation}
L_x^{G}=[M_{G}, L^{G}]
\end{equation}
and
\begin{equation}
L_x^{Q}=[M_{Q}, L^{Q}]
\end{equation}
where the first one is restricted to the surface $ M_1=\{(p,q)\in R^{2N}|<p,q>=0,\
<\Lambda q,q><p,p>+<\Lambda q,p>=0\}$ in the space $R^{2N}$,
respectively produce the finite-dimensional systems:
\begin{eqnarray}
 \  \left\{\begin{array}{l}
q_{x}=\frac{1}{\alpha}(-\Lambda q+(<p,p>-<\Lambda q,q>)p)-p,  \\
p_{x}=\frac{1}{\alpha}(\Lambda p+(<p,p>-<\Lambda q,q>)\Lambda q)+\Lambda q,
  \end{array}
  \right. \label{G}
\end{eqnarray}
and
\begin{eqnarray}
 \  \left\{\begin{array}{l}
q_{x}=\Lambda q+\frac{1}{\beta}<\Lambda q,q>p+\frac{1}{2\beta^2}<p,p><\Lambda q,q>q,\\
p_{x}=-\Lambda p-\frac{1}{\beta}<p,p>\Lambda q-\frac{1}{2\beta^2}<p,p><\Lambda q,q>p.
\end{array}
\right.\label{Q1}
\end{eqnarray}}

Eqs. (\ref{G}) and (\ref{Q1}) turn out to be
the spectral problem studied by Geng
(simply called G-spectral problem) \cite{G1}
\begin{equation}
y_{x}=\left(\begin{array}{cc}
-\lambda u & v-1\\
\lambda (v+1) & \lambda u
 \end{array}\right)y
 \end{equation}
with the constraint condition
\begin{eqnarray*}
u&=&\frac{1}{\alpha}=\frac{1}{\sqrt{(<p,p>-<\Lambda q,q>)^2-4<\Lambda q,p>}},\\
v&=&\frac{<p,p>-<\Lambda q,q>}{\alpha}=\frac{<p,p>-<\Lambda q,q>}{\sqrt{(<p,p>-<\Lambda q,q>)^2-4<\Lambda q,p>}},
\end{eqnarray*}
$\lambda=\lambda_j, \ y=(q_j, p_j)^T$,
and the spectral problem
 proposed by Qiao (simply called Q-spectral problem) \cite{Q3}
\begin{equation}
y_{x}=\left(\begin{array}{cc}
\lambda-\frac{1}{2}uv & u\\
\lambda v & -\lambda+\frac{1}{2}uv
 \end{array}\right)y
 \end{equation}
with the constraint condition
\begin{eqnarray*}
u&=&\frac{<\Lambda q,q>}{\beta}=\frac{<\Lambda q,q>}{1-<q,p>},\\
v&=&\frac{-<p,p>}{\beta}=-\frac{<p,p>}{1-<q,p>},
\end{eqnarray*}
$\lambda=\lambda_j, \ y=(q_j, p_j)^T$, respectively.

 So, Eqs. (\ref{G}) and (\ref{Q1}) are the constrained Geng
 (c-G) flow and the constrained Qiao (c-Q) flow, and they have
the same non-dynamical $r$-matrix:
\begin{equation}
r_{12}(\lambda,\mu)=\frac{2}{\mu-\lambda}P-\frac{2}{\mu}S, \
S=\left(\begin{array}{cccc}
0 & 0 & 0 & 0 \\
0 & 0 & 0 & 0 \\
0 & 1 & 0 & 0 \\
0 & 0 & 0 & 0
\end{array}\right)=\sigma_{-}\otimes\sigma^{+}. \label{rGQ}
\end{equation}

Here, the $r$-matrix $r_{12}(\lambda,\mu)$ can be also chosen as
\begin{equation}
r_{12}(\lambda,\mu)=\frac{2}{\mu-\lambda}P-\frac{2}{\mu}S+I\otimes \tilde{S}.
  \label{rQ}
\end{equation}
Eq. (\ref{rGQ}) is the simplest case: $\tilde{S}=0$ of Eq. (\ref{rQ}).

\vspace{0.65cm}

 We have already seen that the
$r$-matrix $r_{12}(\lambda,\mu)$ satisfying
the fundamental Poisson bracket is not unique
(in fact, infinitely many) and
is usually composed of two parts,
the first one being their main term,
and the second one being the common term $I\otimes \tilde{S}$.
Usually, to prove the integrability we choose their main term
as the simplest $r$-matrix.

\section{An equivalent pair with different  r-matrices}
\setcounter{equation}{0}
This section reveals the following
interesting fact: a pair of constrained systems,
produced by two gauge equivalent spectral problems,
possesses different $r$-matrices.

In 1992, Geng introduced the following spectral problem \cite{G2}
\begin{equation}
\phi_{x}=M\phi, \ M=\left(\begin{array}{cc}
i\la-i\beta uv & u\\
v & -i\la+i\beta uv
 \end{array}\right),\ i^2=-1 \label{GS}
 \end{equation}
where $u$ and $v$ are two scalar potentials, $\la$ is a spectral
parameter and $\beta$ is a constant, and discussed
its evolution equations and Hamiltonian structure.
Eq. (\ref{GS}) is apparently an extension of the AKNS spectral problem
(\ref{AKNS}). Two years later the author considered an extension of the Dirac
spectral problem (\ref{DS}) \cite{Q5}
\begin{equation}
\psi_{x}=\overline{M}\psi, \ \overline{M}=\left(\begin{array}{cc}
-is & \lambda+r+\beta (s^2-r^2)\\
-\la+r-\beta(s^2-r^2) & is
 \end{array}\right), \label{QS}
 \end{equation}
where $r,\ s$ are two potentials,
and obtained a finite dimensional involutive system being not equivalent
to that one in ref. \cite{G2}. But,
the spectral problems (\ref{GS})
and (\ref{QS}) are gauge equivalent via the following transformation \cite{WS}
\begin{equation}
\psi=G\phi,\ G=\left(\begin{array}{cc}
1 & 1\\
i & -i
 \end{array}\right), \label{Gau}
 \end{equation}
$v=i(r-s), \ u=-i(r+s)$. In ref. \cite{WS}, Wadati and Sogo discussed
the gauge transformations of some spectral problems like Eq. \ref{Heis}.

Now, we discuss their $r$-matrices. Let us
consider the following two Lax matrices:{\scriptsize
\begin{eqnarray}
L^{GX}=L^{GX}(\lambda,p,q)=\left(\begin{array}{cc}
 1+2i\beta<p,q> & 0\\
 0 & -1-2i\beta<p,q>
 \end{array}\right)-iL_0, \label{GXLAX}\\
L^{QZ}=L^{QZ}(\lambda,p,q)=\left(\begin{array}{cc}
 0 & \frac{1}{2}-\beta(<p,p>+<q,q>)\\
 -\frac{1}{2}+\beta(<p,p>+<q,q>) & 0
 \end{array}\right)+L_0.\label{QZLAX}
\end{eqnarray}}
Then calculating their determinants leads to the
following Hamiltonian systems
\begin{eqnarray}
 \  \left\{\begin{array}{l}
q_{x}=\frac{\partial H_{GX}}{\partial p}=
\Lambda q+i\beta\frac{<p,p><q,q>}{(1+2i\beta<p,q>)^2}q
            -\frac{<q,q>}{1+2i\beta}p,  \\
p_{x}=-\frac{\partial H_{GX}}{\partial q}=
-\Lambda p-i\beta\frac{<p,p><q,q>}{(1+2i\beta<p,q>)^2}p
+\frac{<p,p>}{1+2i\beta}q,
  \end{array}
  \right. \label{GX}
\end{eqnarray}
and
\begin{eqnarray}
 \  \left\{\begin{array}{l}
q_{x}=\frac{\partial H_{QZ}}{\partial p}=
\Lambda p-\beta\frac{4<p,q>^2+(<p,p>-<q,q>)^2}{(1-2\beta(<p,p>+<q,q>))^2}p
-\frac{2<p,q>q+(<p,p>-<q,q>)p}{1-2\beta(<q,q>+<p,p>)},\\
p_{x}=-\frac{\partial H_{QZ}}{\partial q}=
-\Lambda q+\beta\frac{4<p,q>^2+(<p,p>-<q,q>)^2}{(1-2\beta(<p,p>+<q,q>))^2}q
+\frac{2<p,q>p-(<p,p>-<q,q>)q}{1-2\beta(<q,q>+<p,p>)},
\end{array}
\right.\label{QZ}
\end{eqnarray}
with the Hamiltonian functions
\be
H_{GX}=i<\Lambda q,p>-\frac{<p,p><q,q>}{2(1+2i\beta<p,q>)} \ee
and
\be
H_{QZ}=\frac{1}{2}<\Lambda p,p>+\frac{1}{2}<\La q,q>
-\frac{4<p,q>^2+(<p,p>-<q,q>)^2}{4-8\beta(<p,p>+<q,q>)}. \ee

Obviously, Eqs. (\ref{GX}) and (\ref{QZ}) become Eqs. (\ref{GS}) and
(\ref{QS}) with the constrants
\be
u=-\frac{<q,q>}{1+2i\beta<p,q>}, \  v=\frac{<p,p>}{1+2i\beta<p,q>}, \ee
$\la=\la_j, \  \phi=(q_j,p_j)^T,\ j=1,\ldots,N$; and the constraints
\be
s=\frac{-2i<p,q>}{1-2\beta(<q,q>+<p,p>)}, \  r=\frac{-<p,p>+<q,q>}
{1-2\beta(<q,q>+<p,p>)}, \ee
$\la=\la_j, \  \psi=(q_j,p_j)^T,\ j=1,\ldots,N$, respectively.
Thus, the finite dimensional Hamiltonian systems (\ref{GX}) and (\ref{QZ})
are respectively the constrained flows of the spectral problems
(\ref{GS}) and (\ref{QS}). Since they have Lax matrices (\ref{GXLAX}) and
(\ref{QZLAX}), then the $r$-matrices of (\ref{GX}) and (\ref{QZ})
are respectively:
\be
r_{12}(\la,\mu)=\frac{2}{\mu-\la}P+4i\beta S, \
S=\left(\begin{array}{cccc}
1 & 0 & 0 & 0 \\
0 & 0 & 0 & 0 \\
0 & 0 & 0 & 0 \\
0 & 0 & 0 & 1
\end{array}\right), \label{rGX}\ee
and
\be
r_{12}(\la,\mu)=\frac{2}{\mu-\la}P+2\beta S, \
S=\left(\begin{array}{cccc}
0 & 0 & 0 & -1 \\
0 & 0 & 1 & 0 \\
0 & 1 & 0 & 0 \\
-1 & 0 & 0 & 0
\end{array}\right). \label{rQZ}\ee
which are apparently different.

\section{New integrable systems}
\setcounter{equation}{0}
In this section, three new
integrable systems are generated as the representatives from
our generalized $r$-matrix structure.

 {\bf 1. The first system} is given by case $6$ in section 4. The corresponding
$r$-matrix and involutive systems are respectively
\be
r_{12}(\la,\mu)=\frac{2}{\mu-\la}P+\frac{2}{\mu}S, \
S=\left(\begin{array}{cccc}
1 & 0 & 0 & 0 \\
0 & 0 & 0 & 0 \\
0 & 1 & 0 & 0 \\
0 & 0 & 0 & 1
\end{array}\right) \label{r1}\ee
and
\be
 E^{1}_{j}=2(<p,q>+c)\la_j^{-1}p_jq_j+<q,q>\la_j^{-1}p_{j}^2-\Gamma_j, \
 j=1,\ldots,N, \label{Ej8}
\ee
where $c\in R$.  Thus,
the finite dimesional Hamiltonian systems $(F^1_m)$ defined by
 $F^1_m=\sum_{j=1}^N\la_j^mE^1_j$,
$m=0,\ldots,$ i.e.
\begin{eqnarray}
 F^{1}_{m}&=&2(<p,q>+c)<\la_j^{m-1}p,q>+<q,q><\la_j^{m-1}p,p> \nonumber\\
          & &-\sum_{i+j=m-1}(<\La^i q,q><\La^j p,p>-<\La^i q,p><\La^j p,q>)
            \label{F1}
\end{eqnarray}
are completely integrable. Particularly, with $m=2$ the Hamiltonian system
$(F^1_2)$:
\begin{eqnarray}
 \  \left\{\begin{array}{l}
q_{x}=\frac{\pa F^1_2}{\pa p}=
2c\La q-2<\La q,q>p+4<\La p,q>q+4<p,q>\La q, \\
p_{x}=-\frac{\pa F^1_2}{\pa q}=
-2c\La p+2<p,p>\La q-4<\La p,q>p-4<p,q>\La p,
  \end{array}
  \right. \label{F1H}
\end{eqnarray}
is a new integrable system, which becomes
the following spectral problem
\begin{equation}
\phi_{x}=\left(\begin{array}{cc}
(2c+4v)\la+4u & -2w\\
2s\la & -(2c+4v)\la-4u
 \end{array}\right)\phi \label{SP1}
 \end{equation}
with the constraint conditions $u=<\La p,q>,\ v=<p,q>,\ w=<\La q,q>,
\ s=<p,p>$, and $\la=\la_j,\ \phi=(q_j,p_j)^T, \ j=1,\ldots,N.$
Apparently, the spectral problem (\ref{SP1}) is  new.

\vspace{0.2cm}

  {\bf 2. The second system} is produced by case $(7.1)$ in section 4. The corresponding
$r$-matrix and involutive systems are respectively
\be
r_{12}(\la,\mu)=\frac{2}{\mu-\la}P+\frac{2}{\mu} S, \
S=\left(\begin{array}{cccc}
0 & g'_2 & 0 & 0 \\
f'_2 & 0 & -1 & 0 \\
0 & -1 & 0 & 0 \\
0 & 0 & 0 & 0
\end{array}\right) \label{r2}\ee
and {\small
\be
 E^{2}_{j}=2c\la_j^{-1}p_jq_j+(<q,q>+g_2)\la_j^{-1}p_{j}^2
 +(<p,p>-f_2)\la_j^{-1}q_j^2-\Gamma_j, \ j=1,\ldots,N \label{Ej9}
\ee}
where $c\in R$, $f_2=f_2(<p,q>),\ g_2=g_2(<p,q>)\in C^{\infty}(R)$.
Hence,
the Hamiltonian system $(F^2_2)$ defined by
 $F^2_2=\sum_{j=1}^N\la_j^2E^2_j$, i.e.
{\small \be
 F^{2}_{2}=2c<\La p,q>+2<\La p,q><p,q>+g(<p,q>)<\La p,p>
 -f(<p,q>)<\La q,q>          \label{F2}
\ee}
is completely integrable. Meanwhile the Hamiltonian system
$(F^2_2)$:
{\scriptsize \begin{eqnarray}
q_{x}=\frac{\pa F^2_2}{\pa p}
     &=&2c\La q+2<\La q,p>q+2<p,q>\La q         \nonumber\\
     & &+2g(<p,q>)\La p+<\La p,p>g'(<p,q>)q-<\La q,q>f'(<p,q>)q, \\
p_{x}=-\frac{\pa F^2_2}{\pa q}
     &=&-2c\La p-2<\La q,p>p-2<p,q>\La p \nonumber\\
     & &-<\La p,p>g'(<p,q>)p+2f(<p,q>)\La q-<\La q,q>f'(<p,q>)p,
\end{eqnarray}}
can also related to a new $2\times2$ spectral problem
with some constraint conditions.

\vspace{0.2cm}

  {\bf 3. The third system} is derived by case $(7.4)$ in section 4. The corresponding
$r$-matrix and involutive systems are respectively
\be
r_{12}(\la,\mu)=\frac{2}{\mu-\la}P-\frac{2}{\mu} S, \
S=\left(\begin{array}{cccc}
1 & 0 & 0 & 0 \\
0 & 0 & 0 & 0 \\
0 & 1 & 0 & 0 \\
0 & 0 & c_{-1}' & 1
\end{array}\right) \label{r3}\ee
and
{\small \be
 E^{3}_{j}=2(-<p,q>+c)\la_j^{-1}p_jq_j+<q,q>\la_j^{-1}p_{j}^2
 -c_{-1}\la_j^{-1}q_j^2-\Gamma_j \label{Ej10}
\ee}
where $c\in R$ and $c_{-1}=c_{-1}(<p,q>)\in C^{\infty}(R)$.
The following Hamiltonian system $(F^3_2)$:
{\scriptsize \begin{eqnarray}
 \  \left\{\begin{array}{l}
q_{x}=\frac{\pa F^3_2}{\pa p}=
2c\La q-<\La q,q>(c_{-1}'(<p,q>)q+2p),\\
p_{x}=-\frac{\pa F^3_2}{\pa q}=
-2c\La p+2(c_{-1}+<p,p>)\La q+c_{-1}(<p,q>)<\La q,q>p,
  \end{array}
  \right. \label{F3H}
\end{eqnarray}}
is one of their products, where
\begin{eqnarray}
 F^3_2&=&2c<\La p,q>-<\La q,q>(c_{-1}+<p,p>).
        \label{F3}
\end{eqnarray}
In general, with any $c_{-1}$ Eq. (\ref{F3H}) can't be changed
to a $2\times2$ spectral problem with some constraints.
But with two special $c_{-1}$: $c_{-1}=0$ and $c_{-1}=<p,q>$,
Eq. (\ref{F3H}) can respectively become
the spectral problem \cite{LZ1}
\begin{equation}
\phi_{x}=\left(\begin{array}{cc}
2c\la & -2v\\
2u\la & -2c\la
 \end{array}\right)\phi \label{SP31}
 \end{equation}
with the constraint conditions $u=<p,p>,\ v=<\La q,q>$, and
the spectral problem
\begin{equation}
\phi_{x}=\left(\begin{array}{cc}
2c\la-v & -2v\\
2u\la & -2c\la+v
 \end{array}\right)\phi \label{SP3}
 \end{equation}
with the constraint conditions $u=<p,q+p>,\ v=<\La q,q>$.
Here in Eqs. (\ref{SP31}) and (\ref{SP3})
$\la=\la_j,\ \phi=(q_j,p_j)^T, \ j=1,\ldots,N$ are set.
Eq. (\ref{SP3}) is a new spectral problem.

{\bf Remark 7.1} \ \
We can consider
further new integrable systems induced by Theorem 3.1.

{\bf Remark 7.2} \ \ The above procedure actually give an approach
how to connect an $r$-matrix of finite dimensional system
with a spectral problem, which is closely associated
with integrable NLEEs.

\section{Algebro-geometric solutions}
\setcounter{equation}{0}
The ideal aim for nonlinear differential equations
is of course to obtain their explicit
solution. In this section, we connect the finite dimesional
integrable flows with integrable NLEEs,
and solve them with an explicit form of algebro-geometric
solutions. Here, we take two examples: one being the periodic
or infinite Toda lattice equation,
the other the AKNS equation with the condition
of decay at infinity or periodic boundary.

\subsection{Toda lattice equation}

The Toda hierarchy associated with Eq. (\ref{sp1}) is derived as follows:
\begin{equation}
\left(\begin{array}{l}
u_n\\
v_n
\end{array}\right)_{t_{j}}=JG_j^n, \quad j=0,1,2,\ldots \label{te}
\end{equation}
where {\small $\{G_j^n=J^{-1}KG^n_{j-1}\}_{j=0}^{\infty}$} is the Lenard sequence,
$G^n_{-1}=(\alpha u^{-1}_n,\beta)^T\in Ker J $,
for all
$\alpha=\alpha(t_j),\ \beta=\beta(t_j)\in C^{\infty}(R)$, the two
symmetric operators  $K$, $J$ are
{\small \begin{equation}
K=\left(\begin{array}{lr}
\frac{1}{2}u_n(E-E^{-1})u_n & u_n(E-1)v_n\\
v_n(1-E^{-1})u_n & 2(u_n^2E-E^{-1}u^2_n)
\end{array}\right), \
J=\left(\begin{array}{lr}
0 &  u_n(E-1)\\
(1-E^{-1})u_n & 0
\end{array}\right).
\end{equation}}
In particular, with $j=0$, $\beta=1$ Eq. (\ref{te}) reads as the Toda lattice
\begin{equation}
\dot{u}_n=u_n(v_{n+1}-v_n),\quad \dot{v}_n=2(u_n^2-u_{n-1}^2)\label{te1}
\end{equation}
which can be changed to
\begin{equation}
\ddot{x}_n=2(e^{2(x_{n+1}-x_n)}-e^{2(x_n-x_{n-1})}) \label{te2}
\end{equation}
via the following transformation
\be
u_n=e^{x_{n+1}-x_n},\quad v_n=\dot{x}_n. \label{tran}
\ee

It is easy to prove the following theorem.

 {\bf Theorem 8.1.1}

{\it  1) Let $\hat{G}_n=(\hat{G}^{(1)}_n,\hat{G}^{(2)}_n)^T,\
 \forall \hat{G}^{(1)}_n,\hat{G}^{(2)}_n \in C^{\infty}(R)$. Then
the operator equation
$$
\left[V(\hat{G}_n),L\right]=L_{*}(K\hat{G}_n)-L_{*}(J\hat{G}_n)L
$$
possesses the operator solution
\begin{equation}
V(\hat{G}_n)=-(E^{-1}u_n)\hat{G}_n^{(2)}E^{-1}+
\frac{1}{2}((E^{-1}u_n\hat{G}_n^{(1)})-u_n\hat{G}_n^{(1)})+u_n\hat{G}_n^{(2)}E
\end{equation}
where $\left[\cdot,\cdot \right]$ is the usual commutator; the operator
$L$ is defined by Eq. (\ref{sp1});
 $L_{*}(\xi)=E^{-1}\xi_1+\xi_2+\xi_1E, \ \forall \xi=(\xi_1,\xi_2)^T,
\xi_1, \xi_2\in C^{\infty}(R)$.

    2) Let us choose the special $\hat{G}_n=G^{n}_j,\ j=-1,0,1,...,$ then the Toda
    hierarchy (\ref{te})
has the following Lax representation of operator form
\begin{equation}
L_{t_j}=[W(G^n_j),L],\quad  j=0,1,2...
\end{equation}
where the operator $W(G^n_j)=\sum_{k=0}^{j}V(G_{k-1}^n)L^{j-k}$.}

   Particularly, the standard Toda equation (\ref{te2}) possesses the
 Lax representation of operator form $L_{t}=[W(G^n_0),L]$,
 where the operator $W(G^n_0)=e^{x_{n+1}-x_n}E-e^{x_n-x_{n-1}}E^{-1}$, and
 $u_n,\ v_n$ in $L$ are substituted by (\ref{tran}).

We have shown that the c-Toda flow and the
c-CKdV flow share a common nondynamical $r$-matrix and in particular,
this ensures the integrability of their flows. A calculation of determinant
yields their common $N$-involutive systems
\begin{equation}
 E_{\alpha}=\lambda_{\alpha}p_{\alpha}q_{\alpha}-p_{\alpha}^2-<p,q>q_{\alpha}^2
-\sum_{\beta\not=\alpha,\beta=1}^N \frac{(q_{\alpha}p_{\beta}-p_{\alpha}q_{\beta})^2}
{\lambda_{\alpha}-\lambda_{\beta}},
\alpha=1,\ldots,N
\end{equation}
which are independent and invariant (i.e.
$E_{\alpha}(\la,p,q)=E_{\alpha}(\la,p',q')$).
Apparently, the functions $F_s=\sum^N_{\alpha=1}\la^s_\alpha E_{\alpha},
\ s=0,1,2...$, are given by
{\small \begin{eqnarray}
F_{s}&=&<\Lambda^{s+1}p,q>-<\Lambda^sp,p>-<p,q><\Lambda^sq,q>
   \nonumber \\
     & &-\sum_{j+k=s-1}(<\Lambda^jp,p><\Lambda^kq,q>-<\Lambda^jp,q>
 <\Lambda^kq,p>)
\end{eqnarray}}
and $\{F_m,F_l\}=0$,\ $\forall m,l\in Z^{+}$ which implies the Hamiltonian
systems $(F_s)$ are completely integrable.

    Let $(p_0(t_{s}),q_0(t_{s}))^T$ be a solution of the initial problem
\begin{equation}
\frac{\partial}{\partial t_{s}}\left(\begin{array}{l}
p\\
q
\end{array}\right)=\left(\begin{array}{l}
-\partial F_s/\partial q\\
\partial F_s/\partial p
\end{array}\right),\quad
\left(\begin{array}{l}
p\\
q
\end{array}\right)_{t_{s}=0}=\left(\begin{array}{l}
p_0\\
q_0
\end{array}\right).
\end{equation}
Set
\begin{equation}
\left(\begin{array}{l}
p_n(t_s)\\
q_n(t_s)
\end{array}\right)=H_T^n\left(\begin{array}{l}
p_0(t_s)\\
q_0(t_s)
\end{array}\right)
\end{equation}
where $H_T$ is defined by Eq. (\ref{ctd}). Now, we rewrite
Eq. (\ref{uvn1}) as a map $f: R^{2N} \longrightarrow R^2$ defined by
\be
f:\ \ (p_n,q_n)^T\longmapsto (u_n,v_n)^T. \label{f}
\ee
Then, we have the following theorem.

    {\bf Theorem 8.1.2} \quad{\it
$(u_n(t_s),v_n(t_s))^T=f(p_n(t_s),q_n(t_s))$ satisfies
the Toda hierarchy
\begin{equation}
\frac{d}{d t_{s}}\left(\begin{array}{l}
u_{n}\\
v_{n}
\end{array}\right)=JG^{n}_{s}, \quad s=0,1,\ldots.
\end{equation}}

    Particularly, with $s=0$ the following calculable method
\begin{equation}
\left(\begin{array}{l}
p_0\\
q_0
\end{array}\right)\stackrel{F_0}{\rightarrow}
\left(\begin{array}{l}
p_0(t)\\
q_0(t)
\end{array}\right)\stackrel{H^n}{\rightarrow}\left(\begin{array}{l}
p_n(t)\\
q_n(t)
\end{array}\right)\stackrel{f}{\rightarrow}\left(\begin{array}{l}
u_n(t)\\
v_n(t)
\end{array}\right)
\end{equation}
produces a solution of Toda lattice equation (\ref{te1}).
Thus, the standard Toda equation (\ref{te2}) has the following formal
solution
\be
x_n(t)=\int <q_n(t),q_n(t)>dt.
\ee
We shall concretely give the expression $<q_n(t),q_n(t)>$.

Let us rewrite the element $C_{TC}(\la)$ of Eq. (\ref{CTLax}) as
\be
C_{TC}(\lambda) \equiv -\frac{Q(\lambda)}{K(\lambda)},\quad
K(\lambda)=\prod_{\alpha=1}^{N}(\lambda-\lambda_{\alpha}),\ee
and choose $N$ distinct real zero points $\mu_1,\ldots,\mu_N$
of $Q(\lambda)$. Then, we have
\begin{eqnarray}
Q(\lambda)=\prod_{j=1}^{N}(\lambda-\mu_{j}), \
<q,q>=\sum_{\alpha=1}^{N}\lambda_{\alpha}-\sum_{j=1}^{N}\mu_j. \label{Q}
\end{eqnarray}
Let
\be
\pi_j=A_{TC}(\mu_j),
\ee
then it is easy to prove the following proposition.

    {\bf Proposition 8.1.1} \quad {\it
\begin{equation}
\{\mu_{i}, \mu_{j}\}=\{\pi_{i}, \pi_{j}\}=0,\quad  \{\pi_{j}, \mu_{i}\}=\delta_{ij},
 \quad  i, j=1, 2, ..., N,
\end{equation}
i.e. $\pi_j,\mu_j$ are conjugated, and thus they
are the seperated variables \cite{SK}.}

Write $\det L^{TC}(\la)
=-A_{TC}^2(\la)-B_{TC}(\la)C_{TC}(\la)
=-\frac{1}{4}\la^2-\sum^N_{\alpha=1}\frac{E_{\alpha}}
{\la-\la_{\alpha}}=-\frac{P(\lambda)}{K(\lambda)}$,
where $E_{\alpha}$ is defined by (8.8), and
$P(\lambda)$ is an $N+2$ order polynomial of $\lambda$ whose
first term's coefficient
is $\frac{1}{4}$, then
$ \pi_j^2=\frac{P(\mu_j)}{K(\mu_j)},\ j=1,\ldots,N.$
Now, we choose the generating function
\be
W=\sum_{j=1}^NW_j(\mu_j,\{E_{\alpha}\}_{\alpha=1}^N)=\sum_{j=1}^N
\int_{\mu_j(0)}^{\mu_j(n)}\sqrt{\frac{P(\lambda)}{K(\lambda)}}d \lambda
\ee
where $\mu_j(0)$ is an arbitrary given constant.
Let us view $E_{\alpha} \ (\alpha=1,\ldots,N)$ as actional variables, then angle-coordinates
$Q_{\alpha}$ are chosen as
$$
 Q_{\alpha}=\frac{\partial W}{\partial E_{\alpha}}, \quad \alpha=1,\ldots,N
$$
i.e.
\begin{equation}
Q_{\alpha}=\sum_{k=1}^N \int_{\mu_k(0)}^{\mu_k(n)}\tilde{\omega}_{\alpha}, \quad
\tilde{\omega}_{\alpha}=\frac{\prod^N_{k\not={\alpha},k=1}(\lambda-\lambda_k
)}{2\sqrt{K(\lambda)P(\lambda)}}d\lambda, \quad \alpha=1,\ldots,N.
\end{equation}

    Hence, on the symplectic manifold $(R^{2N},dE_{\alpha}\wedge dQ_{\alpha})$
the Hamiltonian function $F_0=\sum_{\alpha=1}^NE_{\alpha}$ produces a linearized flow
\be
\left\{\begin{array}{l}
\dot{Q}_{\alpha}=\frac{\partial F_0}{\partial E_{\alpha}},\\
\dot{E}_{\alpha}=0,
\end{array}\right.
\ee
thus
\begin{equation}
\left\{\begin{array}{l}
Q_{\alpha}(n)=Q_{\alpha}^0+t+c_{\alpha}n, \quad  c_{\alpha}=\sum_{k=1}^N \int_{\mu_k(n)}^{\mu_k(n+1)}
      \tilde{\omega}_{\alpha},\\
E_{\alpha}(n)=E_{\alpha}(n-1),
\end{array}\right.
\end{equation}
where $c_{\alpha}$ are dependent on actional variables
$\{E_{\alpha}\}^N_{\alpha=1}$, and independent of
$t$; $Q_{\alpha}^0$ is an arbitrary fixed constant.

    Choose a basic system of closed paths $\alpha_i,\beta_i,\ i=1,\ldots,N$
of Riemann surface $\bar{\Gamma}$:\ $\mu^2=P(\lambda)K(\lambda)$ with $N$ handles.
$\tilde{\omega}_j$ ($j=1,\ldots,N$) are exactly $N$ linearly independent holomorphic differentials
 of the first kind
on this Riemann surface $\bar{\Gamma}$.
$\tilde{\omega}_j$ are normalized as $\omega_j=\sum_{l=1}^Nr_{j,l}\tilde{\omega}_l$,
i.e. $\omega_j$ satisfy
$$ \oint_{\alpha_i}\omega_j=\delta_{ij},\quad  \oint_{\beta_i}\omega_j=B_{ij} $$
where $B=(B_{ij})_{N\times N}$ is symmetric and the imaginary part $ImB$ of $B$
is a positive definite matrix.

    By Riemann Theorem \cite{GH} we know: $\mu_k(n)$ satisfies $\sum_{k=1}^N \int
_{\mu_k(0)}^{\mu_k(n)}\omega_j=\phi_j,\
 \phi_j=\phi_j(n,t)\stackrel{\triangle}{=}
\sum_{l=1}^N r_{j,l}(Q_l^0+t+c_ln), \ j=1,\ldots,N$ iff
$\mu_k(n)$ are the zero points of the Riemann-Theta function
   $\tilde{\Theta}(P)=\Theta(A(P)-\phi-K)$
which has exactly $N$ zero points, where $A(P)=
(\int^P_{P_0}\omega_1,\cdots,\int^P_{P_0}\omega_N)^T$, $\phi=\phi(n,t)
=(\phi_1(n,t),\cdots,\phi_N(n,t))^T$, $K\in {\bf C}^N$ is the Riemann constant vector,
$P_0$ is an arbitrary given point on Riemann surface $\bar{\Gamma}$.

    Because of \cite{D}
\be
\frac{1}{2\pi i}\oint_\gamma\lambda d\ln\tilde{\Theta}(P)=C_1(\bar{\Gamma})
\ee
where the constant $C_1(\bar{\Gamma})$ has nothing to do with $\phi$;
$\gamma$ is the boundary of simple connected domain obtained through cutting
the Riemann surface $\bar{\Gamma}$ along closed paths $\alpha_i,\beta_i$. Thus,
we have a key equality
\begin{equation}
\sum_{k=1}^N\mu_k(n)=C_1(\bar{\Gamma})-Res_{\lambda=\infty_1}\lambda d\ln \tilde{\Theta}(P)
-Res_{\lambda=\infty_2}\lambda d\ln \tilde{\Theta}(P) \label{mun}
\end{equation}
where
    $\infty_1:=(0,\sqrt{P(z^{-1})K(z^{-1})}|_{z=0}),\ \infty_2:=
(0,-\sqrt{P(z^{-1})K(z^{-1})}|_{z=0})$.
Through a lengthy careful calculation and combining (\ref{Q}), we obtain
\begin{equation}
<q_n(t),q_n(t)>=\sum_{\alpha=1}^N\lambda_{\alpha}-C_1(\bar{\Gamma})
+\frac{d}{dt}\left(\ln\frac
{\Theta(\phi(n,t)+K+\eta_1)}{\Theta(\phi(n,t)+K+\eta_2)}\right)
\end{equation}
where the $j$-th component of $\eta_i \ (i=1,2)$ is
$\eta_{i,j}=\int^{P_0}_{\infty_i}\omega_j$. By the Riemann surface properties,
 we can also have
$\sum_{l=1}^Nr_{j,l}c_l=\int^{\infty_1}_{\infty_2}\omega_j=\sum_{l=1}^Nr_{j,l}
\int^{\infty_1}_{\infty_2}\tilde{\omega}_l$ which implies
$c_l=\int^{\infty_1}_{\infty_2}\tilde{\omega}_l=\int^{\infty_1}_{P_0}\frac
{\prod_{i\not=l,i=1}^N(\lambda-\lambda_i)}{\sqrt{P(\lambda)K(\lambda)}}d\lambda.
$
So, the standard Toda equation (\ref{te2}) has the following
explicit solution, called {\bf algebro-geometric
solution}
\begin{equation}
x_n(t)=\ln\frac{\Theta(Un+Vt+Z)}{\Theta(U(n+1)+Vt+Z)}+Cn+Rt+const.
\end{equation}
where $U=\hat{R}\hat{C},\ V=\hat{R}\hat{J},\
Z=\hat{R}Q^0+K+\eta_1$ with $\hat{C}=(c_1,\ldots,c_N)^T,\
\hat{J}=(1,\ldots,1)^T,\ Q^0=(Q^0_1,\ldots,Q^0_N)^T$,
$R=\sum^N_{\alpha=1}\lambda_{\alpha}-C_1(\bar{\Gamma})$,
while matrix $\hat{R}
=(r_{j,l})_{N\times N}$ is determined by the relation
$\sum_{l=1}^Nr_{j,l}\oint_{\alpha_i}\tilde{\omega}_l
=\delta_{ij}$, 
and $C$ is certain constant which can be determined by the algebro-geometric
properties on the Riemann surface $\bar{\Gamma}$ \cite{DKV}.
The symmetric matrix $B=(B_{ij})_{N\times N}$
in $\Theta$ function
is determined by
$\sum_{l=1}^Nr_{j,l}\oint_{\beta_i}\tilde{\omega}_l=B_{ij}$.

   Hence, the algebro-geometric solution of Toda lattice
   (\ref{te1}) is
\begin{equation}
\left\{\begin{array}{l}
u_n(t)=e^{x_{n+1}-x_n}=e^{C}\cdot\frac{\Theta^2(U(n+1)+Vt+Z)}{\Theta(U(n+2)+Vt+Z)\Theta(Un+Vt+Z)},\\
v_n(t)=\dot{x}_n=R+\frac{d}{dt}\ln\frac{\Theta(Un+Vt+Z)}{\Theta(U(n+1)+Vt+Z)}.
\end{array}\right.\label{tuv}
\end{equation}
Obviously, the algebro-geometric solution $u_n(t)$ and $v_n(t)$ given by
(\ref{tuv})
are quasi-periodic functions, and they are periodic iff $U=\frac{M}{N}$, where $M$ is a
$N$-dimensional integer column vector. It is easy to see that (\ref{tuv}) is the finite-band
solution of Toda lattice (\ref{te1}) if $\lambda_1,\ldots, \lambda_N$ are chosen
   as the eigenvalues of Toda spectral problem (\ref{sp1}).

\subsection{AKNS equation}
In subsection 5.3 we have shown that the constrained AKNS flow shares a common
$r$-matrix with the constrained Dirac flow, therefore they are integrable.
Now, we derive the algebro-geometric solution for the second order
AKNS equation (\ref{ae1}).

It is well-known that the AKNS hierarchy is given by
\begin{equation}
\left(\begin{array}{l}
u\\
v
\end{array}\right)_{t_{j}}=JG_j, \quad j=0,1,2,\ldots \label{aknse}
\end{equation}
where $\{G_j=J^{-1}KG_{j-1}\}_{j=0}^{\infty}$ is the Lenard sequence,
$G_{-1}=(0,0)^T\in Ker J $,
the two
symmetric operators $K$, $J$ are
\begin{equation}
K=\left(\begin{array}{lr}
2u\pa^{-1}u & \pa-2u\pa^{-1}v\\
\pa -2v\pa^{-1}u & -2v\pa^{-1}v
\end{array}\right), \quad
J=2\left(\begin{array}{lr}
0 &  -1\\
1 & 0
\end{array}\right).
\end{equation}
A representative equation ($j=2$) of (\ref{aknse}) is
\be
u_t=-\frac{1}{2}u_{xx}+u^2v,\  v_t=\frac{1}{2}v_{xx}-v^2u,\ t=t_2. \label{ae1}
\ee

 The independent $N$-involutive system
 of the constrained AKNS flow is expressed by
Eq. (\ref{Ej3}).
Similarly, we consider the following Hamiltonian functions
{\scriptsize \begin{eqnarray}
F_s^{AKNS}&=&\sum^N_{j=1}\la^s_j E_{j}^{AKNS} \nonumber\\
 &=&2<\La^{s}p,q>-\sum_{j+k=s-1}(<\Lambda^jp,p><\Lambda^kq,q>-<\Lambda^jp,q>
 <\Lambda^kq,p>).
\end{eqnarray}}
Let $(p(x,t_{s}),q(x,t_{s}))^T$ be
the involutive solution of the consistent
Hamiltonian canonical equations $(H_{AKNS}),\ (F_s^{AKNS})$.
Then, we have the following theorem.

    {\bf Theorem 8.2.1} \quad{\it
$u=-<q(x,t_j),q(x,t_j)>, \ v=<p(x,t_j),p(x,t_j)>, \ j=0,1,2...,$ satisfy
the higher-order AKNS equations (\ref{aknse}).
Particularly, Eq. (\ref{ae1}) is solved with the following solution:
\be
u=-<q(x,t_2),q(x,t_2)>, \ v=<p(x,t_2),p(x,t_2)>, \label{auv} \ee
where $(p(x,t_{2}),q(x,t_{2}))^T$ is
the involutive solution of the consistent Hamiltonian systems
$(H_{AKNS}),\ (F_2^{AKNS})$.}

In the following procedure we shall express Eq. (\ref{auv}) in
an explicit form of algebro-geometric solution. To do so,
let us rewrite Eq. (\ref{LAKNS}) as follows:
\begin{eqnarray}
L^{AKNS}
=\left(\begin{array}{cc}
   A_{AKNS}(\la) & B_{AKNS}(\la)\\
   C_{AKNS}(\la) & -A_{AKNS}(\la)
 \end{array}\right)
  \label{LAKNS1}
\end{eqnarray}
where
\begin{eqnarray}
   A_{AKNS}(\la)&=&1+\sum_{j=1}^N\frac{1}{\la-\la_j}p_jq_j,\\
   B_{AKNS}(\la)&=&-\sum_{j=1}^N\frac{1}{\la-\la_j}q_j^2,\\
   C_{AKNS}(\la)&=&\sum_{j=1}^N\frac{1}{\la-\la_j}p_j^2.
\end{eqnarray}
$B_{AKNS}(\la),\ C_{AKNS}(\la)$ can be changed to
 the following fractional form:
\begin{eqnarray}
B_{AKNS}(\la) \equiv -\frac{<q,q>Q_B(\la)}{K(\la)},\quad
C_{AKNS}(\la) \equiv \frac{<p,p>Q_C(\la)}{K(\la)}
\end{eqnarray}
where
\begin{eqnarray*}
<q,q>Q_B(\la)&=&\sum_{j=1}^Nq_j^2\prod^N_{k=1,k\not=j}(\la-\la_k),\\
<p,p>Q_C(\la)&=&\sum_{j=1}^Np_j^2\prod^N_{k=1,k\not=j}(\la-\la_k),\\
K(\la)&=&\prod_{j=1}^{N}(\lambda-\lambda_{j}).
\end{eqnarray*}
Respectively choosing $N-1$ distinct real zero points $\mu_1^B,\ldots,\mu_{N-1}^B$
and $\mu_1^C,\ldots,\mu_{N-1}^C$ of $Q_B(\la)$ and $Q_C(\la)$ leads to
\begin{eqnarray}
\sum_{k=1}^{N-1}\mu_k^B \ = \ A_1-\frac{<\La q,q>}{<q,q>}, \ \
\sum_{k=1}^{N-1}\mu_k^C \ = \ A_1-\frac{<\La p,p>}{<p,p>},  \label{mbc11} \\
(A_1-\sum_{k=1}^{N-1}\mu_k^B)^2-\sum_{k=1}^{N-1}(\mu_k^B)^2 \ = \
          2A_2-A_1^2+2\frac{<\La^2 q,q>}{<q,q>}, \label{mb12}\\
(A_1-\sum_{k=1}^{N-1}\mu_k^C)^2-\sum_{k=1}^{N-1}(\mu_k^C)^2 \ = \
         2A_2-A_1^2+2\frac{<\La^2 p,p>}{<p,p>}, \label{mc12}
\end{eqnarray}
where $ A_1=\sum_{j=1}^{N}\la_{j}, \ A_2=
\sum_{k,j=1,j<k}^{N}\la_{j}\la_k$ are two constants.
One hand, $u_x=-2<q,q_x>=-2<q,\frac{\pa H_{AKNS}}{\pa p}>=-2<\La q,q>-2uc_0(t)$,
here $c_0(t)$ is an arbitrarily fixed function of $t$.
Thus from Eq. (\ref{mb12}) we have
\begin{eqnarray} \frac{\pa}{\pa x}\ln u=2A_1-2\sum_{k=1}^{N-1}\mu_k^B-2c_0(t).\end{eqnarray}
On the other hand, $u_{t_2}=-2<q,q_{t_2}>=-2<q,\frac{\pa F_2^{AKNS}}{\pa p}>
=-2<\La^2 q,q>$. This is combined with Eq. (\ref{mb12}) to give the equality
\begin{eqnarray}
\frac{\pa}{\pa t_2}\ln u=(A_1-\sum_{k=1}^{N-1}\mu_k^B)^2-
\sum_{k=1}^{N-1}(\mu_k^B)^2-2A_2+A_1^2. \end{eqnarray}
So, we obtain
\begin{eqnarray}
u(x,t)&=&u(x_0,t_0)\exp(\int_{t_0}^t\left[(A_1-\sum_{k=1}^{N-1}\mu_k^B)^2-
            \sum_{k=1}^{N-1}(\mu_k^B)^2-2A_2+A_1^2\right]dt \nonumber \\
      & &\ \ \ \ \ \ \ \ \ \ \ \ +\int_{x_0}^x
             \left[2A_1-2\sum_{k=1}^{N-1}\mu_k^B-2c_0(t)\right]dx), \
      t=t_2, \label{u}
\end{eqnarray}
where $x_0,\ t_0$ are two fixed initial values. Similarly, $v(x,t)$ has the
following representation
\begin{eqnarray}
v(x,t)&=&v(x_0,t_0)\exp(-\int_{t_0}^t\left[(A_1-\sum_{k=1}^{N-1}\mu_k^C)^2-
               \sum_{k=1}^{N-1}(\mu_k^C)^2-2A_2+A_1^2\right]dt \nonumber \\
      & &\ \ \ \ \ \ \ \ \ \ \ \ -\int_{x_0}^x\left[
      2A_1-2\sum_{k=1}^{N-1}\mu_k^C-2c_0(t)\right]dx), \
      t=t_2. \label{v}
\end{eqnarray}
Since Eqs. (\ref{u}) and (\ref{v}) solves nonlinear
equation (\ref{ae1}), then in order to obtain their explicit form
it needs calculating the four key expressions
$\sum_{k=1}^{N-1}(\mu_k^J)^k$, $J=B, C; \ k=1,2$.
For that purpose, we follow the approach in the case
of Toda lattice equation. For the present two set of Darboux coordinates
$\mu^J_j$, $J=B,C$; $j=1,...,N-1$,
we have the following key equalities
like Eq. (\ref{mun})
\begin{eqnarray}
\sum_{j=1}^{N-1}(\mu_j^J)^k=C_k(\Gamma)-\sum_{s=1}^2Res_{\la=\infty_s}\la^k
 d\ln \Theta(A(P)-\phi-K_J), \\ J=B, C; \ k=1,...,N-1, \nonumber
\end{eqnarray}
where $C_k(\Gamma)$ is a constant \cite{Qiao2,Zhou1} only determined by the
compact Riemann surface $\Gamma$ ($genus=N-1$):
$\mu^2=P_{AKNS}(\la)K(\la)$, $P_{AKNS}(\la)=K(\la)+\sum_{j=1}^NE^{AKNS}_j
\prod^N_{k\not=j,k=1}\\(\la-\la_k)$; {\small $\infty_1=(0,\sqrt{P_{AKNS}(z^{-1})
K(z^{-1})}|_{z=0}), \ \infty_2=
(0,-\sqrt{P_{AKNS}(z^{-1})K(z^{-1})}|_{z=0})$}; $A(P)=\int^P_{P_0}\omega$
is an Abel map in which $P_0$ is an arbitrarily fixed point on $\Gamma$,
$\omega=\\(\omega_1,...,\omega_{N-1})^T, \omega_j=\sum_{l=1}^{N-1}
r_{j,l}\tilde{\omega}_l=\sum_{l=1}^{N-1}r_{j,l}
\frac{\prod^N_{k\not=l,k=1}(\lambda-\lambda_{k})}
{2\sqrt{K(\la)P_{AKNS}(\la)}}d\la$ is a normalized holomorphic
differential form, and $r_{j,l}$ is the normalized factor;
The $j$-th component $\phi_j(x,t)$ of $N-1$ dimensional vector $\phi$ equals to
$\sum_{l=1}^{N-1}r_{j,l}(Q_l^0+\frac{1}{2}\la_lx+\frac{1}{2}\la_l^2t+C_l(t)+
\tilde{C}_l(x))$ with the arbitrary constant $Q_l^0$ and functions
$C_l(t), \ \tilde{C}_l(x)\in C^{\infty}(R)$; $K_B,\ K_C\in {\bf C}^{N-1}$ are
the two Riemann constant vectors respectively associated with the Darboux
coordinates $\mu_j^B,\ \mu_j^C$; Riemann-Theta function \cite{MD}
$\Theta(\xi)$ is defined on Riemann surface $\Gamma$.

A lengthy calculation of Residue at $\infty_s, \ s=1,2$ for $k=1,2$ yields
\begin{eqnarray}
\sum_{j=1}^{N-1}\mu_j^J&=&C_1(\Gamma)-\frac{\pa}{\pa x}\ln\frac{\Theta_1^J}
       {\Theta_2^J},\\
\sum_{j=1}^{N-1}(\mu_j^J)^2&=&C_2(\Gamma)+\frac{\pa}{\pa t}\ln
       \frac{\Theta_1^J}{\Theta_2^J}-\frac{\pa^2}{\pa x^2}\ln\Theta_1^J\Theta_2^J,
\end{eqnarray}
where $\Theta_s^J=\Theta(\phi+K_J+\eta_s),\ J=B, C, \ \eta_{s,j}=\int^{P_0}
_{\infty_s}\omega_j,  s=1,2$ is the $j$-th component of the $N-1$ dimensional
vector $\eta_s$.

  Substituting the above equalities into (\ref{u}) and (\ref{v}),
  and sorting them, we
obtain the explicit solution of nonlinear equation (\ref{ae1}):
\begin{eqnarray}
u(x,t)&=&u(x_0,t_0)e^{a(t-t_0)+2(b-c_0(t))(x-x_0)}\frac{\Theta_1^B}
       {\Theta_2^B}\vert_{t=t_0}(\frac{\Theta_2^B}{\Theta_1^B})^2\vert_{x=x_0} \nonumber\\
     & &\times \frac{\Theta_1^B}{\Theta_2^B}\exp(\int_{t_0}^t\left[
         \frac{\pa^2}{\pa x^2}\ln\Theta_1^B\Theta_2^B+(b+\frac{\pa}{\pa x}\ln
         \frac{\Theta_1^B}{\Theta_2^B})^2\right]dt),\label{aeu}\\
v(x,t)&=&v(x_0,t_0)e^{-a(t-t_0)-2(b-c_0(t))(x-x_0)}\frac{\Theta_2^C}
       {\Theta_1^C}\vert_{t=t_0}(\frac{\Theta_1^C}{\Theta_2^C})^2\vert_{x=x_0} \nonumber \\
      & &\times \frac{\Theta_2^C}{\Theta_1^C}\exp(\int_{t_0}^t\left[
  \frac{\pa^2}{\pa x^2}\ln\Theta_2^C\Theta_1^C+(b+\frac{\pa}{\pa x}\ln
  \frac{\Theta_2^C}{\Theta_1^C})^2\right]dt), \label{aev}
\end{eqnarray}
where $a=A_1^2-C_2(\Gamma)-2A_2,\ b=A_1-C_1(\Gamma)$ are two constants,
$c_0(t)\in C^{\infty}(R)$ is an arbitrarily given function of $t$, and
$x_0, \ t_0$ are the initial values. Therefore, we have the following
theorem.

{\bf Theorem 8.2.2} \quad{\it
The AKNS equation (\ref{ae1})
has the explicit solution (\ref{aeu}) and (\ref{aev})
given by the form of Riemann-Theta function, which
is called the {\bf algebro-geometric solution}.}

  An analogous calculational process will lead to the algebro-geometric
  solution of the higher-order AKNS equation (\ref{aknse}).

\section{Conclusions and problems}

   In finite dimensional case,
   Lax matrix is enough to provide many important integrable properties
   like $r$-marix, Hamiltonian,
integrability, Darboux coordinates, and even later algebro-geometric solution.
Therefore we specially stress to use
Lax matrix instead of Lax pair in finite dimensional case.

   The generalized $r$-matrix structure is given
   to emphasize the classification and united sketch of finite
   dimensional integrable systems.
   We have already seen that only is there one concrete $r$-matrix
   structure, then the corresponding Hamiltonian flows
are surely integrable and even in some cases the associated
spectral problems are new.

  In the paper, we develope our generalized structure
   to solve some integrable equations with algebro-geometric solution.
 This is an extension of nonlinearization methods \cite{C1}.
It is found that this procedure
can be also applied into other integrable NLEEs
\cite{Zhou1, Zhang1, Du1}.
In this sense,
we successfully realize a procedure from finite
dimensional flows to infinite dimensional systems
when we have some constrained or restricted relation between them.
Of course, there are still other methods to solve integrable NLEEs.
Recently, Deift, Its and Zhou \cite{DIZ,DZ}
obtained the $\Theta$-function solutions of some
integrable NLEEs like the KdV, MKdV, nonlinear Schr$\ddot{\rm o}$dinger
equation by using Riemann-Hilbert asymptotic method.
All these methods are still under the development.

It should be pointed out that our
procedure is carried in the symplectic space
$(R^{2N}, \ dp\wedge dq)$ (i.e. corresponding to
the Bargmann constraint).
How about the case restricted on a subsymplectic manifold in the
space $R^{2N}$ (i.e. corresponding to the C. Neumann constraint)? This is
a difficult problem. Although $r$-matrix works out
\cite{ZQ}, and there is no answer about the algebro-geometric
solution up to now.

   From section 5, we know that
the c-Toda (or r-Toda) flow and
the c-CKdV (or r-Toda) flow share the same
$r$-matrix as well as the common Lax matrix and involutive conserved integrals
in the whole space $R^{2N}$ (or on certain symplectic submanifolds in $R^{2N}$).
Therefore
a further conjecture is: whether any finite dimensional continuous
Hamiltonian flow can be associated with a finite dimensional discrete
symplectic map such that they share a common Lax matrix? If it is right,
then the discrete integrable systems will be mostly enlarged.

\vspace{0.2cm}
\section*{\Large \bf Acknowledgments}

The author would like to express his sincere thanks to
Prof. Gu Chaohao and Prof. Hu Hesheng for their enthusiastic
instructions and helps, and thanks Prof. Cao Cewen, Prof. Liu Zhangju and
Prof. Strampp for their helpful discussions and advices.
He thanks very much for Dr. Holm for his fruitful discussion.
He is also very grateful to the Fachbereich 17 of the University Kassel,
Germany, in particular to Prof. Strampp and Prof. Varnhorn for their warm
invitation and hospitality, and to the referees
for their precious suggestions. 

This work has been supported by the Alexander von Humboldt
Foundation, Germany; the Chinese National Basic Reseach Project
``Nonlinear Science'', the Special Grant of Chinese National Excellent
Ph. D. Thesis, and the Doctoral Programme Foundation of the Institution of
High Education, China.

                      \end{document}